\documentclass[preprint,aip,jcp,a4paper,floatfix]{revtex4-1}

\usepackage{amssymb,amsmath}
\usepackage{graphicx}

\begin{document}

\title{Nonequilibrium fluctuation-dissipation relations
for one- and two-particle correlation functions in
steady-state quantum transport}

\author{H. Ness}
\affiliation{Department of Physics, School of Natural and Mathematical Sciences,
King's College London, Strand, London WC2R 2LS, UK}
\email{herve.ness@kcl.ac.uk}
\affiliation{Department of Physics, University of York, Heslington, York YO10 5DD,UK}
\affiliation{European Theoretical Spectroscopy Facility (ETSF)}

\author{L. K. Dash}
\affiliation{Department of Physics and Astronomy, University College London, Gower Street, London WC1E 6BT, UK}
\affiliation{European Theoretical Spectroscopy Facility (ETSF)}

\begin{abstract}
  We study the non-equilibrium (NE) fluctuation-dissipation (FD) relations in the context of 
  quantum thermoelectric transport through a two-terminal nanodevice in the steady-state.
  The FD relations for the one- and two-particle correlation functions are derived for a
  model of the central region consisting of a single electron level.
  Explicit expressions for the FD relations of the Green's functions (one-particle correlations)
  are provided. 
  The FD relations for the current-current and charge-charge (two-particle)
  correlations are calculated numerically.
  We use self-consistent NE Green's functions calculations to treat the system in the absence and 
  in the presence of interaction (electron-phonon) in the central region.
  We show that, for this model, there is no single universal FD theorem for the NE steady state. 
  There are different FD relations for each different class of problems.
  We find that the FD relations for the one-particle correlation function are strongly dependent on 
  both the NE conditions and the interactions, while the FD relations of the current-current correlation 
  function are much less dependent on the interaction.
  The latter property suggests interesting applications for single-molecule and other nanoscale transport 
  experiments.
\end{abstract}
 

\maketitle

\newlength{\mycolumnwidth}
\setlength{\mycolumnwidth}{85mm}

\section{Introduction}
\label{sec:intro}

Quantum systems can be driven far from equilibrium either by time-dependent perturbation
or by coupling to reservoirs at different chemical potentials or temperatures.
In the latter case, the system is said to be open because it can exchange particles and/or
energy with the reservoirs, Hence particle and/or energy currents flow 
throughout the system. 
Such processes take place in different contexts, ranging from nanoscale 
thermo-electric conductors to bio-chemical reactions.
The recent developments in modern techniques of microscopic manipulation and nanotechnologies
enable us to build functional nanoscale systems, for example, electronic nanodevices or 
molecular motors. 
Transport properties and fluctuations in such systems can nowadays be experimentally 
resolved \cite{Blanter:2000,Forster:2008,Cuevas:book2010,Widawsky:2012}.

At equilibrium, small fluctuations in a system satisfy a universal relation
known as the fluctuation-dissipation (FD) theorem \cite{Kubo:1957,Kubo:1966}.
The FD relation connects spontaneous fluctuations to the linear response of
the system and holds for both the classical and quantum cases. 

The search for similar relations for systems driven far from equilibrium has been an 
active area of research for many decades. A major breakthrough had taken place with 
the discovery of exact fluctuation relations, which hold for classical systems at 
non-equilibrium (NE) \cite{Bochkov:1977,Bochkov:1979,Jarzynski:1997,Jarzynski:2004,Seifert:2010}.
The derivation of fluctuation theorems for quantum systems has also been 
considered in Refs.~[\onlinecite{Yukawa:2000,Kurchan:2000,Andrieux:2008,Campisi:2011b}], 
either in the context of quantum heat conduction or electron full counting
statistics \cite{Jarzynski:2004,Saito:2007,Andrieux:2008,Gelin:2008,Andrieux:2009,Esposito:2009,
Talkner:2009,Flindt:2010,Campisi:2011a,Safi:2011,Seifert:2012}
in open quantum systems.
Another route to study NE FD relations is to find effective local thermodynamical variables 
(temperature, chemical potential) dependent on the NE conditions, but entering the conventional 
equilibrium FD relations \cite{Arrachea:2005,Caso:2010,Caso:2012}.

On one hand, we know that, at equilibrium, the universal expression of the FD theorem 
involves the thermodynamical variables of the system, i.e. the
temperature and the chemical potential. On the other hand, we know that the steady state
regime can be seen as an effective equilibrium state \cite{Hershfield:1993,Ness:2013,Ness:2014}.
It is therefore mostly interesting to know if there exist extensions of the FD theorem
to the NE steady state of an open quantum system connected to reservoirs at their own 
(but different) equilibrium. For instance, we can ask ourselves if there is a unique functional 
form of the FD theorem which depends only on the different thermodynamical variables of the
reservoirs, i.e. their different temperatures and chemical potentials? 

Motivated by understanding such NE properties at the nanoscale and their potential use 
in practical nanodevices, we study, in this paper, generalisations of FD relations to 
NE conditions in the presence of both charge and heat transport in the steady state regime.
We consider a two-terminal nanodevice where the central region consists of a single
electron level in the presence (or absence) of interaction.

For this model, we show that there is no unique universal expression for the FD relations 
in a NE steady state. 
The FD relations depend on the thermodynamical properties of the reservoirs and on
the corresponding forces (gradients) that drive the open quantum system out of equilibrium, 
and on the interaction present in that system.

In particular, we focus on the NE FD relationships for one-particle correlation functions, 
i.e. the electron Green's functions (GFs), and for two-particle correlation functions, 
i.e. the current-current (JJ) and charge-charge (CC) correlation and response functions.
We extend the concept of the Kubo-Martin-Schwinger (KMS) and FD relations to the 
NE steady state regime. 
The expressions for the KMS and FD relations for our model open quantum system connected 
to two reservoirs
(in the presence of a finite applied bias and a temperature gradient) are explicitly derived. 
Furthermore we also consider the many-body (MB) effects on such relations by using a model of 
an interacting electron-phonon system.

We show that the FD relation for the one-particle correlation functions is strongly dependent 
on both the NE conditions and the interaction between particles. While the FD relations for the
two-particle correlation functions are much less dependent on the interaction.
We also briefly discuss how such FD relations could be used to obtain information about the 
properties of the system from measurements. 

To present our results, we have chosen the following organisation of the paper: 
the main analytical and numerical results are given in the main body of the paper while 
technical aspects are provided in the Appendices. 
We provide typical numerical results in the main body of the paper and give additional 
results for different parameters of the model (representing different physical conditions) 
in the Appendices. These results are quantitatively different from each other, however 
they all confirm the trends of the FD relations.

In Sec.~\ref{sec:equiFD}, we recall known relations for the FD theorem at equilibrium.
In Sec.~\ref{sec:NESS}, we focus on the NE steady state regime
for a system consisting of a finite size central region (a single electron level) 
connected to two semi-infinite reservoirs at their own equilibrium.
The expressions for the KMS and FD relations 
for the one-particle correlation functions are derived in Sec.~\ref{sec:1Pcorrel},
for both the non-interacting case (Sec.~\ref{sec:non-inter}), and the interacting case
(Sec.~\ref{sec:interac}).
The two-particle correlation functions are derived in Sec.~\ref{sec:2Pcorrel}, and the
corresponding FD relations are obtained from numerical calculations.
In Sec.~\ref{sec:noninter_case}, we provide results for the non-interacting system, 
for which we discuss the concept of effective equilibrium with an effective 
temperature (Sec.~\ref{sec:Teff_noint}).
The effects of the interaction in the central region are shown in Sec.~\ref{sec:sssm_case}
for a model of local electron-phonon coupling. 
We present results for the off-resonant (Sec.~\ref{sec:interac_offres}) transport regime
(results for resonant transport region are given in Appendix ~\ref{app:interac_res}).
The effects of the strength of the coupling to the leads (of the electron-phonon coupling)
are shown in Sec.~\ref{sec:interac_leadscoupling} (Appendix ~\ref{app:interac_strength}).
We finally discuss potential connections of our results with experiments in Sec. \ref{sec:discuss}
and conclude in Sec. \ref{sec:ccl}.

\section{Equilibrium fluctuation-dissipation relation}
\label{sec:equiFD}

We first recall the important relations obtained from equilibrium
statistical mechanics \cite{Note1}.
At equilibrium, the FD theorem arises from the fact that the time evolution operator
$e^{-iHt}$ bears a strong formal similarity to the weighting factor $e^{-\beta H}$
that occurs in statistical averages by identifying $t \equiv -i \beta$ (with the
usual definition $\beta=1/kT$). 
The key relation is that, for any two operators $A$ and $B$, one has
\begin{equation}
\label{eq:KMS}
\langle A(t-i\beta) B(t') \rangle = \langle B(t') A(t) \rangle \ .
\end{equation}

We define the following correlation functions  
\begin{equation}
\label{eq:XAB}
X^>_{AB} = \langle A(1) B(2) \rangle \ \ \ \ {\rm and} \ \ \ \
X^<_{AB} = \mp \langle B(2) A(1) \rangle \ , 
\end{equation}
with the minus (plus) sign for $A,B$ being 
fermion (boson) operators. The integer $1,2$ represents a composite index
for space and time coordinates $(x_i,t_i)$ (or other convenient index for
the space coordinate if one works on a lattice or with localised basis sets).

At equilibrium, these quantities depend only on the time difference,
and after Fourier transform, one can write the FD theorem for the
correlations $X^\lessgtr_{AB}(\omega)$ as follows:
\begin{equation}
\label{eq:equiFDT}
\begin{split}
X^>_{AB}+X^<_{AB} 
= 
\left[ \frac{r_{AB}(\omega)+1}{r_{AB}(\omega)-1} \right] 
\left( X^>_{AB} - X^<_{AB} \right) \ ,
\end{split}
\end{equation}
with the ratio $r_{AB}(\omega)$ obtained from the equilibrium KMS 
relation \cite{Kubo:1957,Kubo:1966,Kadanoff:1962,Stefanucci:2013}
\begin{equation}
\label{eq:KMSw}
r_{AB}(\omega)=\frac{X^>_{AB}(\omega)}{X^<_{AB}(\omega)} = \mp e^{\beta\bar\omega} \ .
\end{equation}
The minus sign is for fermion operators (plus sign for boson operators)
and $\bar\omega=\omega-\mu^{\rm eq}$ for grand-canonical ensemble
averages with the equilibrium chemical potential $\mu^{\rm eq}$ 
($\bar\omega=\omega$ for the canonical ensemble).

For boson operators $A$ and $B$, the usual relation between commutator and
anticommutator is obtained from Eq.(\ref{eq:equiFDT}):
\begin{equation}
\label{eq:equiFDT_boson}
\begin{split}
\langle \left\{ A, B \right\} \rangle_\omega
= 
\coth \left( {\beta\bar\omega}/{2} \right) 
\langle \left[ A, B \right] \rangle_\omega \ ,
\end{split}
\end{equation}
with $\langle X \rangle_\omega$ being the Fourier transform
of $\langle X(t-t') \rangle$.

For fermion operators with $A=\Psi$ and $B=\Psi^\dag$ being the electron
annihilation and creation operators respectively, the correlation functions 
$X^\lessgtr_{AB}$ are the electron GFs  $X^\lessgtr_{AB}= i G^\lessgtr(\omega)$.
We now recall the usual definitions:
\begin{equation}
\label{eq:defGF}
\begin{split}
G^K     & =  G^>+G^<=i\langle [\Psi , \Psi^\dag] \rangle_\omega \ , \\
G^r-G^a & =  G^>-G^< = i\langle \{\Psi , \Psi^\dag\} \rangle_\omega \ , 
\end{split}
\end{equation}
from which one can recover the equilibrium FD relation:
\begin{equation}
\label{eq:equiFDT_GF}
G^K(\omega)
= 
\tanh \left( \beta\bar\omega/2 \right) 
\left[ G^r(\omega) - G^a(\omega) \right]  \ .
\end{equation}

Furthermore, using the equilibrium KMS ratio $G^>(\omega)/G^<(\omega)= - e^{\beta\bar\omega}$, one
obtains another well known relation 
\begin{equation}
G^< = G^< \frac{G^> - G^<}{G^> - G^<} = - f^{\rm eq} (G^> - G^<) = - f^{\rm eq} (G^r - G^a) \ ,
\end{equation}
with $f^{\rm eq}(\omega)=[1-G^>/G^<]^{-1}=[1+e^{\beta\bar\omega}]^{-1}$ being the
equilibrium Fermi distribution.

All these known equilibrium relations are important to remember for comparison with their
NE counterparts that we develop in the next sections.

\section{Non-equilibrium steady state fluctuation-dissipation relations}
\label{sec:NESS}

\subsection{The quantum transport set-up}
\label{sec:NEtransp}

In this paper, we consider the FD relations of one- and two-particle correlation functions
in the context of non-equilibrium steady state quantum transport. 
We focuss on a system consisting of a central region connected to two electrodes
(modelled by non-interacting Fermi seas).
The central region $C$ may contain interaction characterized by a self-energy 
$\Sigma_{\rm int}$ in the NE GF formalism \cite{Dash:2010,Dash:2011}.
The left ($L$) and right ($R$) electrodes are at their own equilibrium, with a Fermi 
distribution $f_\alpha(\omega)$ defined by their respective 
chemical potentials $\mu_\alpha$ and temperatures $T_\alpha$ ($\alpha=L,R$).
This a typical transport set-up to measure (thermo)electric properties of quantum
dots or single-molecule nanoscale junctions 
\cite{Widawsky:2012,DiVentra:book2008,Cuevas:book2010,Haug:1996}.

The expression we derived in the following sections are obtained for a central
region consisting of a single electron level (i.e. the single impurity model). They
could be generalised to other models with several electronic levels by using the
matrix form of the so-called non-equilibrium distribution function \cite{Ness:2014}. 
We study the simplest possible model system which nonetheless contains
the relevant physics of the transport properties of the molecular
junction \cite{Frederiksen:2007,Lee:2009,Ness:2011,Ness:2012,Ness:2014}.
 Furthermore the specific model used for the electrodes does not need
to be specified, as long as the leads can be described by an embedding
self-energy $\Sigma_{\alpha}(\omega)$ in the electron GF of the central region.

Our results for the FD relations are general with respect to both the leads 
$\Sigma_{\alpha}(\omega)$ and the interaction self-energies $\Sigma_{\rm int}$,
in the same sense that the GFs have a general expression with respect to these
self-energies.

\subsection{The one-particle correlation functions}
\label{sec:1Pcorrel}

The one-particle correlation functions are the electron GFs defined in the central region. 
As shown in Sec.~\ref{sec:equiFD}, they are obtained from two correlation functions:
\begin{equation}
\label{eq:GF}
\begin{split}
G^<(t,t') & = -i \langle d^\dag(t') d(t) \rangle \ , \\ 
G^>(t,t') & =  i \langle d(t) d^\dag(t') \rangle \ , 
\end{split}
\end{equation}
where $d^\dag$ ($d$) creates (annihilates) an electron in the central region and
$\langle\dots\rangle$ is the average over the NE ensemble \cite{Ness:2013}.
In the NE conditions, the GFs are defined on the Keldysh time-loop contour
(Appendix \ref{app:sec:NEGF}).
In the steady state, all quantities depend only on the time difference
$X(t,t')=X(t-t')$ and can be Fourier transformed into an single-energy 
representation $X(\omega)$.

\subsubsection{The non-interacting case}
\label{sec:non-inter}

In the absence of interaction in the central region, one can use the
properties of the non-interacting lesser and greater GFs $G_0^\lessgtr$:
\begin{equation}
\label{eq:GF_G0}
G_0^\lessgtr(\omega)=G_0^r(\omega) \left[
\Sigma^\lessgtr_L(\omega)+\Sigma^\lessgtr_R(\omega) \right] G_0^a(\omega),
\end{equation}
to show that they follow the pseudo-equilibrium relations \cite{Hershfield:1991}:
\begin{equation}
\label{eq:GF_f0NE}
\begin{split}
G_0^<(\omega) & = -  f_0^{\rm NE}(\omega)    \left[ G_0^r(\omega) - G_0^a(\omega) \right] \\
G_0^>(\omega) & = - (f_0^{\rm NE}(\omega)-1) \left[ G_0^r(\omega) - G_0^a(\omega) \right], 
\end{split}
\end{equation}
where 
\begin{equation}
\label{eq:deff0NE}
f_0^{\rm NE}(\omega)=\frac{\Gamma_L(\omega)f_L(\omega)+\Gamma_R(\omega) f_R(\omega)}{\Gamma_L(\omega)+\Gamma_R(\omega)} 
\end{equation}
is the NE distribution function of the central region in the absence of interaction, and
$\Gamma_\alpha(\omega)=i(\Sigma_\alpha^>(\omega)-\Sigma_\alpha^<(\omega))$
is the spectral function of the lead $\alpha=L,R$.

We can then determine the FD ratio (FDR) from Eq.~(\ref{eq:equiFDT})
as follows:
\begin{equation}
\label{eq:FDratioG0}
{\rm FDR}[G_0]=\frac{G_0^>+G_0^<}{G_0^>-G_0^<}=1-2f_0^{\rm NE}(\omega) \ .
\end{equation}
As expected, ${\rm FDR}[G_0]=\tanh \left( \beta\bar\omega/2 \right)$
at equilibrium. The derivation is provided in Appendix \ref{app:FDRnoint}.
In this Appendix we also derived explicit expressions for the KMS and FDR
for systems at a unique temperature, and with symmetric or asymmetric coupling to the leads.
These results have been already derived, in other forms, in the literature. 

However we can extent these results to the cases where there are both a temperature gradient 
and an applied bias (i.e. each lead at a different temperature and a different chemical
potential).
We find that the NE FD ratio for the symmetric non-interacting case is given by:
\begin{equation}
\label{eq:FDratioG0_withgradT}
\begin{split}
&{\rm FDR}[G_0](V,\nabla T) =
\left[ 2\sinh \bar\beta\bar\omega \right] /  \left[  2\cosh \bar\beta\bar\omega \right. \\   
& + \left( e^{-\beta_L V/2} + e^{\beta_R V/2}\right) \cosh (\Delta\beta\bar\omega/2)         \\
&\left.  + \left( e^{-\beta_L V/2} - e^{\beta_R V/2}\right) \sinh (\Delta\beta\bar\omega/2)  \right]  ,
\end{split}
\end{equation}
with the averaged inverse temperature given by $\bar\beta=(\beta_L+\beta_R)/2$ and 
the gradient of inverse temperature given by $\Delta\beta=\beta_L-\beta_R$.

We can now show an interesting property of similarity between applied bias 
(at fixed unique temperature) and temperature gradient (at fixed bias).
For example, at zero bias $V=0$, Eq.~(\ref{eq:FDratioG0_withgradT}) becomes
\begin{equation}
\label{eq:FDratioG0_V0withgradT}
\begin{split}
{\rm FDR}[G_0](V=0,\nabla T) = 
\frac{ \sinh \bar\beta\bar\omega } 
{ \cosh \bar\beta\bar\omega  + \cosh (\Delta\beta\bar\omega/2)} .
\end{split}
\end{equation}
By comparison Eq.~(\ref{eq:FDratioG0_V0withgradT}) with the FDR obtained at finite bias
$V\ne 0$ and zero temperature gradient ($T_L=T_R$), see Eq.~(\ref{eq:FDratioG0sym}), 
we can notice that
the gradient $\Delta\beta$ plays a similar role as the 
gradient of chemical potential $\Delta\mu=\mu_L-\mu_R=V$ \cite{Note2}.
The important difference between Eq.~(\ref{eq:FDratioG0_V0withgradT}) and Eq.~(\ref{eq:FDratioG0sym})
is that, in the presence of a gradient of temperature between
the leads, the central region is at an effective temperature $T_{\rm eff}$ since
the FD ratio in Eq.~(\ref{eq:FDratioG0_V0withgradT}) is defined from $\bar\beta$.
The effective temperature $T_{\rm eff}$ is
defined from $\bar\beta=1/kT_{\rm eff}$ as $T_{\rm eff}=2 T_L T_R / (T_L+T_R)$.
The concept of local effective temperature, in the presence of an applied bias, 
is examined in further detail in Section \ref{sec:Teff_noint}.

\subsubsection{The interacting case}
\label{sec:interac}

In the presence of an interaction described by the self-energy 
$\Sigma_{\rm int}(\omega)$ in the central region, 
we use again the properties of the NE GF $G^\lessgtr$ to
find that
\begin{equation}
\label{eq:KMSratioG}
\begin{split}
\frac{G^<}{G^>} 
= & \frac{G^r \Sigma^< G^a}{G^r \Sigma^> G^a}
                  =\frac{\Sigma^<_{L+R}+\Sigma^<_{\rm int}}{\Sigma^>_{L+R}+\Sigma^>_{\rm int}} \\
= & \frac{f_0^{\rm NE} - i \Sigma^<_{\rm int}/\Gamma_{L+R}}{f_0^{\rm NE}-1 - i \Sigma^>_{\rm int}/\Gamma_{L+R}} \ .
\end{split}
\end{equation}
From this ratio, we define a NE distribution function $f^{\rm NE}(\omega)=[1-G^>/G^<]^{-1}$
[\onlinecite{Ness:2013}].
The NE distribution enters into the relation between the different GFs in a similar form 
as for the non-interacting case:
\begin{equation}
\label{eq:GF_fNE}
\begin{split}
G^<(\omega) & = -  f^{\rm NE}(\omega)    \left( G^r - G^a \right)(\omega) \\
G^>(\omega) & = - (f^{\rm NE}(\omega)-1) \left( G^r - G^a \right)(\omega) .
\end{split}
\end{equation}
 The NE distribution $f^{\rm NE}(\omega)$ can be expressed as follows:
\begin{equation}
\label{eq:fNE}
f^{\rm NE}(\omega) = \frac{f_0^{\rm NE}(\omega) - i \Sigma^<_{\rm int}(\omega)/\Gamma_{L+R}(\omega)}
{1+i (\Sigma^>_{\rm int}-\Sigma^<_{\rm int} ) / \Gamma_{L+R} } \ ,
\end{equation}
where $\Gamma_{L+R}(\omega)=\Gamma_L(\omega)+\Gamma_R(\omega)$.

In the most general cases, we have shown in Refs.~[\onlinecite{Ness:2013,Ness:2014}] 
that the full NE distribution $f^{\rm NE}(\omega)$ differs 
from the non-interaction distribution $f_0^{\rm NE}(\omega)$.

We can now derive the FD ratio for the interacting GFs as follows:
\begin{equation}
\label{eq:FDratioGdef}
{\rm FDR}[G]=\frac{ G^>(\omega) + G^<(\omega)}{ G^>(\omega) - G^<(\omega)} = 1 - 2 f^{\rm NE}(\omega) \ .
\end{equation}

This expression is similar to the non-interacting case. However,
from Eqs.~(\ref{eq:KMSratioG}) and (\ref{eq:fNE}), we obtain after some algebraic manipulation
a FD ratio which is clearly different from the non-interacting case:
\begin{equation}
\label{eq:FDratioG}
{\rm FDR}[G]=\frac{ {\rm FDR}[G_0] + i ( \Sigma^>_{\rm int} + \Sigma^<_{\rm int} ) / \Gamma_{L+R} }
                      { 1 + i ( \Sigma^>_{\rm int} - \Sigma^<_{\rm int} ) / \Gamma_{L+R}                  } \ .
\end{equation}

The FD ratio in Eq.~(\ref{eq:FDratioG}) depends on both the NE conditions 
($\Delta\mu$ and $\Delta T$ via ${\rm FDR}[G_0]$)
and on the MB effects via the interaction self-energies
(which are themselves dependent on the NE conditions).

We have now derived the main analytical results for the NE extension of the 
FD relations for the one-particle correlation functions (for our model system).
The NE FD relations, Eq.~(\ref{eq:FDratioG0_withgradT}) and Eq.~(\ref{eq:FDratioG}), 
are less universal than their equilibrium counterparts, in the sense that they depend on both 
the set-up that drives the system out of equilibrium, i.e. a two-terminal configuration with 
a applied finite bias $V$ and a temperature gradient $\Delta\beta$, and on the NE interaction (when present).
However the NE FD relations are universal, with respect to the interaction, in the same sense 
that the GFs have universal expressions with respect to the self-energies.

\subsection{The two-particle correlation functions}
\label{sec:2Pcorrel}

The two-particle correlation functions we consider involve
products of two pairs of electron creation-annihilation operators. Such pairs
can be local or non-local, i.e. diagonal or off-diagonal elements 
in terms of the regions $\alpha=L,C$ or $R$. The local pairs correspond to the
the charge-charge CC correlation function, and the non-local pairs to the 
current-current JJ correlation function.

We now recall the definitions \cite{Blanter:2000} for the fluctuation correlation 
function $S^X_{\alpha\beta}$ for a quantity $X_\alpha$ (the charge or current in 
the region $\alpha=L,C$ or $R$):
\begin{equation}
\label{Snoise}
S^X_{\alpha\beta}(t,t')=\frac{1}{2}
\langle\left\{ \delta X_\alpha(t),\delta X_{\beta}(t') \right\}\rangle \ ,
\end{equation}
where $\delta X_\alpha(t)= X_\alpha(t)- \langle X_\alpha \rangle$.
The definition of the corresponding response function is:
\begin{equation}
\label{R}
R^X_{\alpha\beta}(t,t')=\langle\left[ X_\alpha(t),X_{\beta}(t') \right]\rangle \ .
\end{equation}

For the current flowing at the $\alpha=L,R$ contact, we have $X_\alpha(t)=J_\alpha(t)$. 
The charge current is expressed as follows \cite{Ness:2011,Ness:2012b}: 
\begin{equation}
\label{eq:current}
\langle J_\alpha(t)\rangle = {e}/{\hbar}\ {\rm Tr}_\alpha \left[ (\Sigma G)^<(t,t) - (G \Sigma)^<(t,t) \right].
\end{equation}

For the charge density in the central region, we take $\alpha=\beta=C$ and $X_C(t)=n_C(t)$ with
\begin{equation}
\label{eq:charge}
\langle n_C(t)\rangle = {e}\ {\rm Tr}_C [ -i G^<(t,t) ] .
\end{equation}
The trace ${\rm Tr}_x$ is taken on the electronic degrees of freedom of the region $x=L,C$ or $R$.

In the steady state, all quantities are dependent only on the time difference $\tau=t-t'$, i.e.
$X(t,t')=X(t-t'$) and after
Fourier transform we obtain all quantities with a single-energy argument $X(\omega)$.
The general expressions for the fluctuation and response functions are:
\begin{equation}
\label{SRab_w}
\begin{split}
2S^X_{\alpha\beta}(\omega)  & = 
\langle x_\alpha x_\beta \rangle _\omega^- + \langle x_\beta x_\alpha \rangle _\omega^+  \\
R^X_{\alpha\beta}(\omega)   & = 
\langle x_\alpha x_\beta \rangle _\omega^- - \langle x_\beta x_\alpha \rangle _\omega^+  \ ,
\end{split}
\end{equation}
with
$x_\alpha= j_\alpha$ for the current-current correlation function, and 
$x_\alpha=n_C$ for the charge-charge correlation function.
In Appendix \ref{app:2Pcorrel} we explain in detail how to calculate the two-particle 
correlation functions and in Appendix \ref{app:2Pcorrel_JJ} we provide the detailed 
expressions of the current-current correlation functions.

For practical applications, it is more useful to consider the total noise 
$S^J(\omega)$ and response function $R^J(\omega)$ obtained from the symmetrized
current $J=(J_L-J_R)/2$. The total noise is obtained from the following linear combination
$S^J(\omega)=(S^J_{LL}-S^J_{LR}-S^J_{RL}+S^J_{RR})/4$ (similarly for $R^J(\omega)$).
The equivalent of Eq.~(\ref{SRab_w}) for  $S^J(\omega)$ and $R^J(\omega)$ is then
given by:
\begin{equation}
\label{SRtot_w}
\begin{split}
\left\{ \begin{array}{c} 2S^J(\omega) \\ R^J(\omega) \end{array} \right\}
= 
\langle j j \rangle _\omega^- \pm \langle j j \rangle _\omega^+ .
\end{split}
\end{equation}
The explicit expression for $\langle j j \rangle _\omega^\pm$ is given in Appendix \ref{app:2Pcorrel_JJ}.

We define the NE FD ratio for the JJ correlation functions as:
\begin{equation}
\label{eq:FDR_JJ}
{\rm FDR[JJ]} = 2S^J(\omega) / R^J(\omega) \ .
\end{equation}
At equilibrium, we have the usual FD relation:
\begin{equation}
\label{eq:equiFD_SR}
2S^J(\omega) = \coth \left( {\beta\omega}/{2} \right) R^J(\omega) \ ,
\end{equation}
as the current operator is a boson-like operator \cite{Note3}.

For the charge-charge CC fluctuation $S^N(\omega)$ and response $R^N(\omega)$ functions, we have
\begin{equation}
\label{eq:SRCC_w}
\begin{split}
\left\{ \begin{array}{c} 2S^N(\omega) \\ R^N(\omega) \end{array} \right\}
=  
\langle n_C n_C \rangle _\omega^- \pm \langle n_C n_C \rangle _\omega^+ \ ,
\end{split}
\end{equation}
with
\begin{equation}
\label{eq:NNCC_w}
\langle n_C n_C \rangle _\omega^{\mp}= {e^2} \int \frac{{\rm d}u}{2\pi}\  {\rm Tr}_C [G^>(u) G^<(u\mp\omega) ] \ .
\end{equation}
The FD ratio for the CC correlation functions is defined as:
\begin{equation}
\label{eq:FDR_CC}
{\rm FDR[CC]} = 2S^N(\omega) / R^N(\omega) \ .
\end{equation}
At equilibrium the charge-charge correlation and response functions follow the FD relation 
Eq.~(\ref{eq:equiFD_SR}) since the charge operator is also a boson-like operator \cite{Note4}.

We have now all the tools to present numerical calculations for the NE FDR of the JJ and CC 
correlation functions and compare such a FDR with the FDR of the GFs.
In the figures of the next sections, we actually represent the inverse of the FDR 
$1/{\rm FDR}=R^X(\omega)/2S^X(\omega)$. 
This permits us to avoid the divergence of the $\coth$-like function 
at $\omega=0$ and it allows for a direct comparison with the NE FD ratio of the GFs 
(which behaves as $\tanh {\beta\omega}/{2}$ at equilibrium).

\subsection{Breakdown of the equilibrium relations}
\label{sec:Yoperator}

As shown above, the KMS and FD relations do not hold generally in the NE 
conditions, even if the steady state could be seen as a pseudo-equilibrium 
state \cite{Hershfield:1993,Ness:2013}.
 
In order to understand more quantitatively the origin of such a breakdown, 
we consider the reformulation of NE 
quantum statistical mechanics made by Hershfield \cite{Hershfield:1993}.
In this work, a time-independent NE density matrix 
$\rho^{\rm NE} \equiv e^{-\beta (H-Y)}$ is derived. It incorporates both the 
NE and MB effects \cite{Hershfield:1993,Han:2007,Han:2010,Han:2012}.

The $Y$ operator is constructed by an iterative scheme from the equation
of motion (in the interaction representation) of an initial $Y_0$ operator.
In the case of a two-terminal device, the initial operator is $Y_0=\mu_L N_L + \mu_R N_L$
with the left and right chemical potentials $\mu_{L,R}$ and particle number operators $N_{L,R}$
of the two reservoirs.
The key relation leading the FD theorems becomes with the use of $\rho^{\rm NE}$:
\begin{equation}
\label{eq:NEKMS}
\langle A(t-i\beta) B(t') \rangle = \langle e^{-\beta Y } B(t') e^{\beta Y} A(t) \rangle \ ,
\end{equation}
which is different from the equilibrium relation Eq.~(\ref{eq:KMS}).
The usual equilibrium FD relations break down at NE because additional
contributions arise from the expansion
\begin{equation}
\begin{split}
e^{-\beta Y } B e^{\beta Y} = B - [\beta Y, B] + [\beta Y, [\beta Y, B] ] / 2! \\
- [\beta Y, [\beta Y, [\beta Y, B] ] ]/ 3! + ... \nonumber
\end{split}
\end{equation}

Indeed, any operator $B$ (electron annihilation/creation, charge current, charge density)
does not necessarily commute with the NE operator $Y$.
Hence NE FD relations depend on the full series of commutators $[\beta Y, B]$.
In that sense, these FD relations will be less universal than the equilibrium FD theorem, since they 
depend on the nature of the operator $B$ and on the NE operator $Y$ which includes both the NE and MB 
effects.

\section{Numerical application: the non-interacting case}
\label{sec:noninter_case}

In the absence of interaction, the Hamiltonian for the central region $C$ is simply 
given by $  H_C   = \varepsilon_0 d^\dagger d $
where $d^\dagger$ ($d$) creates (annihilates) an
electron in the level $\varepsilon_0$.

We choose to model the left and right electrodes by one-dimensional
tight-binding chains with hopping integrals $t_{0L}$ and
$t_{0R}$ to the central region. 
The corresponding (retarded) lead self-energy is
$\Sigma^r_\alpha(\omega)=t_{0\alpha}^2/\beta_\alpha e^{{\rm i} k_\alpha}$
with the dispersion relation 
$\omega=\varepsilon_\alpha+2\beta_\alpha \cos k_\alpha(\omega)$.
With such a choice, the lead self-energy is energy dependent and go beyond the wideband limit.

Furthermore, we model the fraction of potential drops at the contacts 
by $\mu_L=\mu^{\rm eq}+\eta_V V$ and $\mu_R=\mu^{\rm eq}-(1-\eta_V) V$,
with an approximation for the $\eta_V$ factor taken from 
[\onlinecite{Datta:1997,Dash:2012}]: $\eta_V=t_{0R}/(t_{0L}+t_{0R})$.

Full NE GF calculations \cite{Dash:2010,Dash:2011} have been performed (with $\Sigma_{\rm int}=0$)
for a wide range of the different parameters.
We consider also different possible regimes, including
symmetric ($t_{0L}=t_{0R}$) and asymmetric ($t_{0L}\ne t_{0R}$)
coupling to the leads, different strengths of coupling $t_{0\alpha}$,
different biases $V$ with symmetric and asymmetric potential drops at the contacts,
different transport regimes (off-resonant $\varepsilon_0 \gg \mu^{\rm eq}$, and
resonant  $\varepsilon_0 \sim \mu^{\rm eq}$),  
and different temperatures.

\begin{figure}
  \center
  \includegraphics[clip,width=\mycolumnwidth,height=55mm]{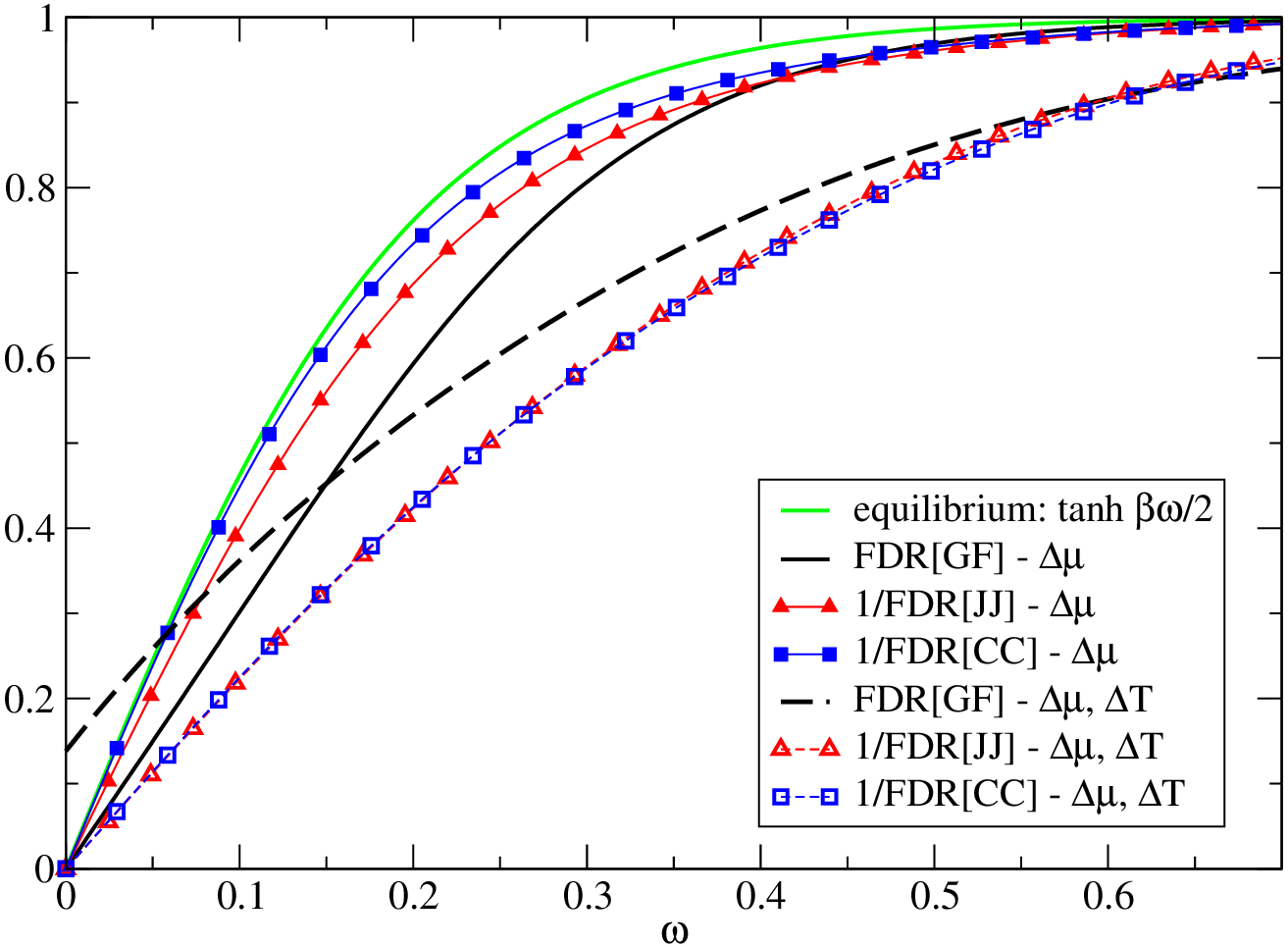}
  \caption{(color online)
FD ratio of the GF Eq.~(\ref{eq:FDratioG0_withgradT}), 
and inverse FDR of the JJ and CC correlation functions, Eqs.(\ref{eq:FDR_JJ})-(\ref{eq:FDR_CC}), 
for the non-interacting case.
Calculations are performed for the off-resonant regime ($\varepsilon_0=0.45$) and symmetric coupling to the leads.
The equilibrium ($V = 0.0, \Delta T =0.0$), electric transport ($\Delta\mu = 0.30, \Delta T =0.0$) and 
thermoelectric transport ($\Delta\mu = 0.30, \Delta T =0.1$) regimes are shown. 
At equilibrium, the FD ratio is given by $\tanh \beta\omega/2$.
The other parameters  $t_{0\alpha}=0.19$, $T_R=T_{\rm ph}=0.1$, $T_L=T_R+\Delta T$, 
$\mu_{L,R}=\mu^{\rm eq}\pm V/2$, $\varepsilon_\alpha=0, \beta_\alpha=2$.}
  \label{fig:1}
\end{figure}

\subsection{FD ratio and NE distribution}
\label{sec:typicalres_noint}

\begin{figure}
  \center
  \includegraphics[clip,width=\mycolumnwidth,height=55mm]{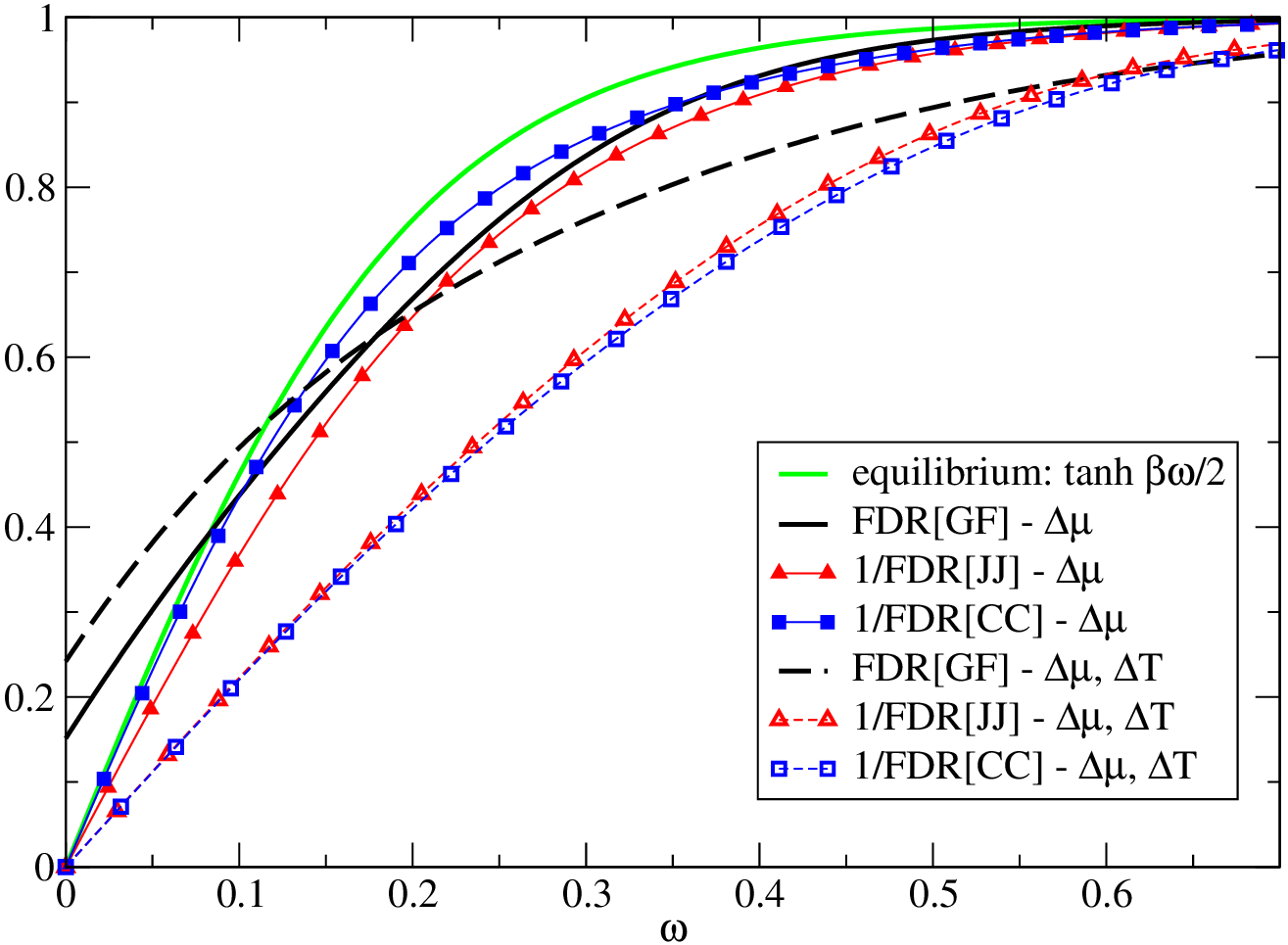}
  \caption{(color online)
FD ratio of the GF Eq.~(\ref{eq:FDratioG0_withgradT}), 
and inverse FDR of the JJ and CC correlation functions, Eqs.(\ref{eq:FDR_JJ})-(\ref{eq:FDR_CC}), 
for the non-interacting case.
Calculations are performed for the off-resonant regime ($\varepsilon_0=0.45$) 
and asymmetric coupling $t_{0L} \ne t_{0R}$.
The equilibrium ($V = 0.0, \Delta T =0.0$), electric transport ($\Delta\mu = 0.30, \Delta T =0.0$) 
and thermoelectric transport ($\Delta\mu= 0.30, \Delta T =0.1$)
regimes are shown. 
The other parameters  $t_{0L}=0.19$, $t_{0R}=0.285$, $\eta_V=0.6$, $T_R=T_{\rm ph}=0.1$, $T_L=T_R+\Delta T$, 
$\mu_L=\mu^{\rm eq}+\eta_V V$, $\mu_R=\mu^{\rm eq}-(1-\eta_V) V$,
$\varepsilon_\alpha=0, \beta_\alpha=2$.}
  \label{fig:2}
\end{figure}

Figures \ref{fig:1} and \ref{fig:2} show typical results for the different FD ratios
of the one- and two-particle correlation functions.
Figure \ref{fig:1} corresponds to the case of a symmetric coupling to the leads,
associated with symmetric potential drops at the contacts, while Figure \ref{fig:2} 
corresponds to a typical asymmetric coupling case.

We can see strong deviations from the equilibrium FD ratio in the presence of the NE
conditions ($\Delta\mu\ne 0$ and $\Delta T \ne 0$). 
The presence of a temperature gradient on top of the applied bias increases the deviation 
from the equilibrium behaviour.

Crudely speaking, the shape of the FDR (inverse FDR for the two-particle correlations) in the presence
of an applied bias $\Delta\mu\ne 0$ (with $\Delta T = 0$) looks like the equilibrium FDR with a different
(smaller) slop at the origin, i.e. with a larger effective temperature.
It seems like applying a bias corresponds to an increase of the local temperature.

\begin{figure}
  \center
  \includegraphics[clip,width=\mycolumnwidth,height=55mm]{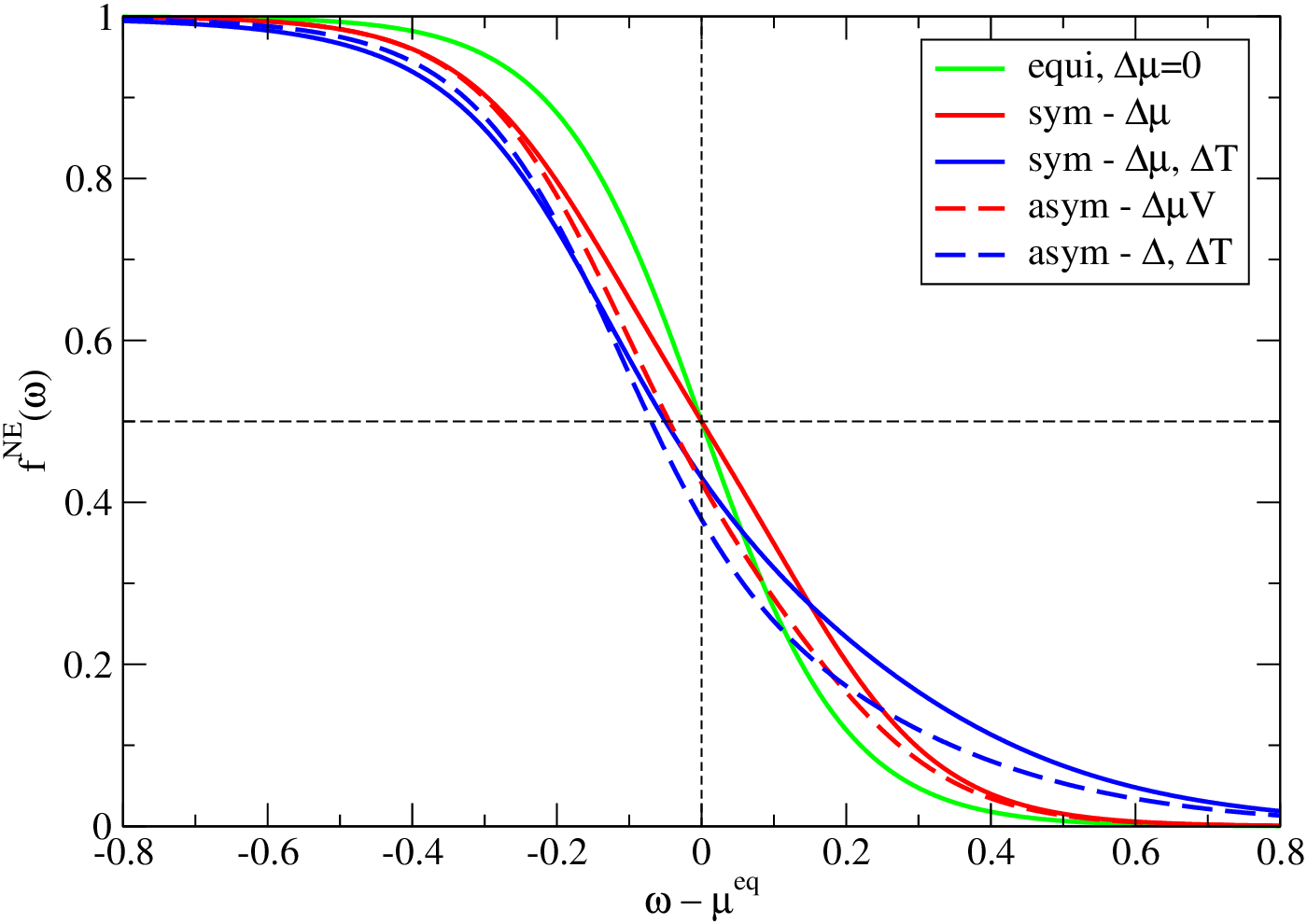}
  \caption{(color online)
NE distribution function $f_0^{\rm NE}$ for the non-interacting case. Calculations are
done for the off-resonant regime, with symmetric and asymmetric coupling to the leads as 
show in figures \ref{fig:1} and \ref{fig:2}. The energy reference is taken with respect 
to the equilibrium chemical potential $\mu^{\rm eq}$.}
  \label{fig:3}
\end{figure}

Furthermore one can see a different behaviour of the FD ratio of
the GFs in comparison to the FDR of the JJ and CC correlations.
Indeed, the limit FDR$\rightarrow 0$ for $\omega\rightarrow 0$ is obtained
for the two-particle correlations in any conditions.
For the one-particle correlation, this limit is obtained only for certain conditions. 
Such a behaviour is related to the intrinsic symmetry of the NE distribution function
$f_0^{\rm NE}$ versus the energy reference (i.e. the chemical potential at equilibrium $\mu^{\rm eq}$)
as shown in Figure \ref{fig:3}.

For the Fermi distribution at equilibrium, there is a symmetry point 
such as $f(-x)=1-f(x)$ where $x$ is the energy with
respect to the equilibrium chemical potential $x=\omega-\mu^{\rm eq}$ (see fig. \ref{fig:3}).
Such a symmetry relation is also valid for the non-interaction NE distribution $f_0^{\rm NE}$ 
when the coupling to the leads is symmetric ($\Gamma_L=\Gamma_R$), with a symmetric
potential drop $\eta_V=1/2$ and a uniform temperature throughout the system ($T_L=T_R$).
As soon as some form of asymmetry is introduced (i.e. asymmetric coupling to the leads 
and/or asymmetric potential drop $\eta_V \ne 0$, and/or temperature gradient $\Delta T \ne 0$),
the relation $f(-x)=1-f(x)$ does not hold any more, and $f(x=0)\ne 0.5$.
Hence ${\rm FDR}[\omega\rightarrow 0] \ne 0$, since the FDR of the GFs is directly related 
to the NE distribution.

Finally, it should be noted that the FDR of the two-particle correlations has always the
limit FDR$\rightarrow 0$ when $\omega\rightarrow 0$, as for the equilibrium FDR.
This comes from the fact that the FDR of the JJ and CC correlations is not directly 
related to the NE distribution, in opposition to the FDR[GF].

\subsection{Effective equilibrium approximation}
\label{sec:Teff_noint}

In the linear response regime (small applied bias), it has been show that the FD ratio 
could be approximated by an effective equilibrium FDR calculated with an effective temperature 
($\beta_{\rm eff}$) which differ from the thermodynamical temperature ($\beta_L$ or $\beta_R$) 
\cite{Kreuzer:1981,Caso:2010,Caso:2012}.
The concept of a single, but effective, temperature in the central region is reasonable
for a small system (i.e. single impurity model) \cite{Lebowitz:1959,Thingna:2013}. 
However in larger systems, one generally observes a temperature profile (not a unique effective 
temperature) in the central region \cite{Dubi:2009}. Such a temperature profile cannot in
general be described by an effective equilibrium \cite{Dubi:2009}.

In the following, we show numerically to which extent the approximation of an effective single
temperature is valid for the JJ correlations in our single impurity model.
For that, we fit the inverse ratio 1/FDR[JJ] onto an effective equilibrium ratio of
the type $\tanh  \beta_{\rm eff}\omega/2$. 
The effective inverse temperature $\beta_{\rm eff}$ is calculated from the derivative 
of 1/FDR[JJ] versus $\omega$ taken in the limit $\omega\rightarrow 0$.

\begin{figure}
  \center
  \includegraphics[clip,width=\mycolumnwidth,height=55mm]{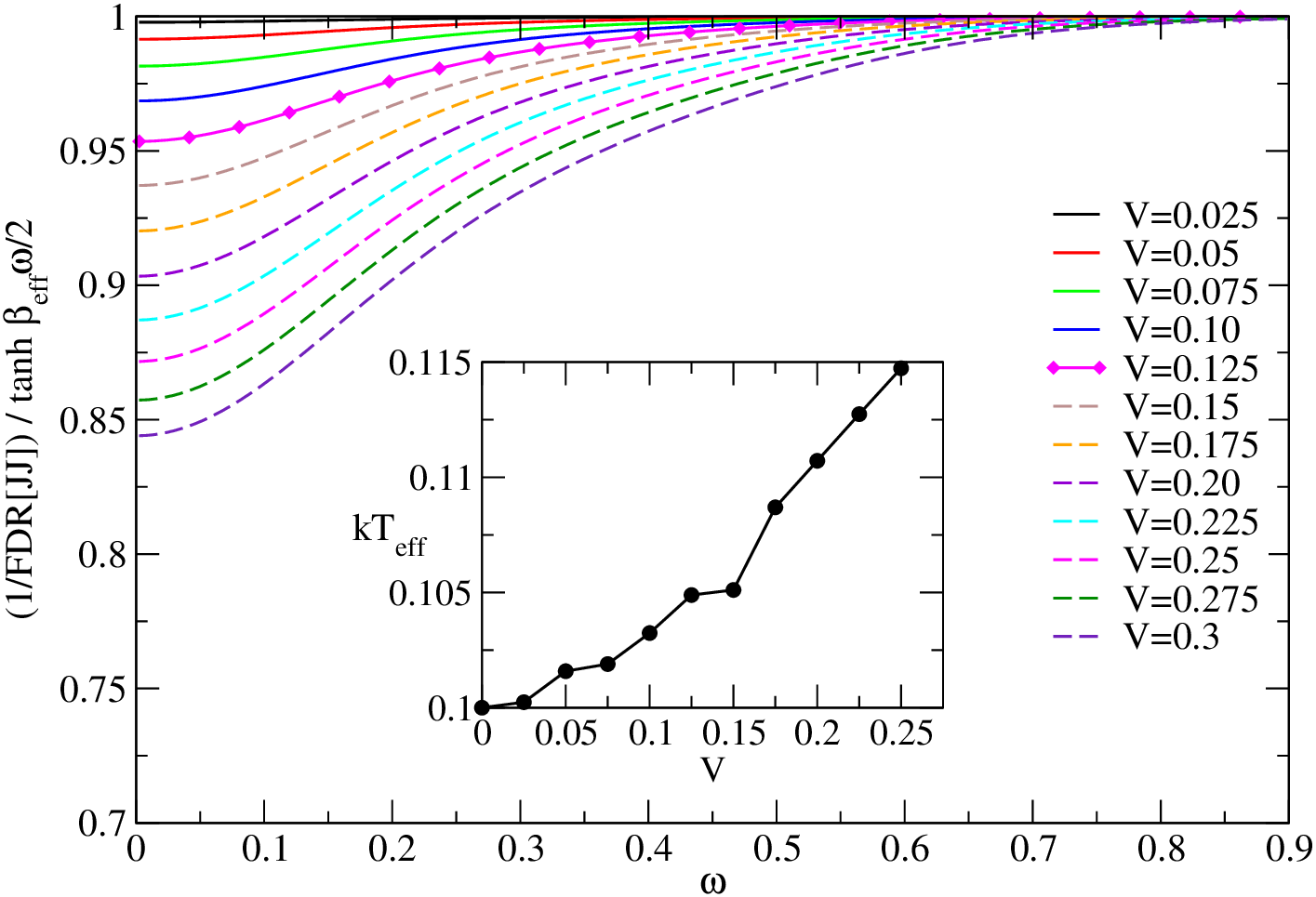}
  \caption{(color online)
Inverse ratio 1/FDR[JJ] normalised by an effective equilibrium FDR
$\tanh \beta_{\rm eff}\omega/2$ for the off-resonant case (with symmetric
coupling to the leads). 
The dependence of the effective temperature $kT_{\rm eff}=1/\beta_{\rm eff}$ versus
applied bias $V$ in shown in the inset.
A quasi linear dependence of $kT_{\rm eff}$ on the applied bias $V$ is obtained
in the linear regime (here $V < 0.15$).
Within this regime, the fits of 1/FDR[JJ] onto $\tanh \beta_{\rm eff}\omega/2$ are 
within a 5\% error bar. The parameters are as in figure \ref{fig:1}.}
  \label{fig:4}
\end{figure}

Figure \ref{fig:4} show the inverse ratio 1/FDR[JJ] divided by the corresponding
effective equilibrium form $\tanh  \beta_{\rm eff}\omega/2$, for a different applied 
biases and for the off-resonant transport regime with symmetric coupling to the lead.
The dependence of the effective temperature $kT_{\rm eff}=1/\beta_{\rm eff}$ versus
the applied bias is also shown in the inset of Figure \ref{fig:4}.

We can observe a quasi linear dependence of the effective temperature $kT_{\rm eff}$ 
upon the driving force $\Delta\mu=V$, for small biases.
For the set of parameters used, we get a good representation of 1/FDR[JJ] by the 
expression $\tanh \beta_{\rm eff}\omega/2$ for biases $V < 0.15$. This
corresponds to the linear regime, for which the fits stay within a 5\% error bar. 
For larger applied bias, the fit onto an effective equilibrium FDR is not appropriate.

It is interesting to note that the effective temperature behaves as 
$T_{\rm eff}=T(1+\alpha V)$ in the linear regime. Such a behaviour is similar to the
results obtained analytically in Ref.~[\onlinecite{Caso:2010}] for an electronic system 
driven out of equilibrium by pumping, i.e. by a local a.c. field in the absence of 
applied bias.

\section{Numerical application: the interacting electron-phonon case}
\label{sec:sssm_case}

We now provide numerical calculations of the NE FDR for a specific choice 
of interaction in the central region.
We consider a model with electron-phonon interaction \cite{Dash:2010,Dash:2011}.
This is a minimal model which contains the essential physics for studying 
inelastic transport properties of single-molecule junctions, as we have shown in 
Refs.~[\onlinecite{Dash:2010,Dash:2011,Dash:2012,Ness:2010,Ness:2012}].

The Hamiltonian for the central region is now
\begin{equation}
\label{eq:H_central}
\begin{split}
  H_C 
  = \varepsilon_0 d^\dagger d + \omega_0 a^\dagger a +
  \gamma_0 (a^\dagger + a) d^\dagger d,
\end{split}
\end{equation}
where $d^\dagger$ ($d$) creates (annihilates) an
electron in the level $\varepsilon_0$ which
is coupled to a vibration mode 
of energy $\omega_0$ via the coupling constant $\gamma_0$.
The operators $a^\dagger$ ($a$) creates (annihilates) a quantum of
vibration in the mode $\omega_0$.
The many-body (MB) electron-phonon interaction self-energies $\Sigma_{\rm int}$ 
are treated at the Hartree-Fock level (first order diagrams in term of the interaction).
Self-consistent calculations provide a partial resummation of the 
diagrams to all orders in the corresponding NE GFs \cite{Dash:2010,Dash:2011}.

We have performed calculations for the same wide range of parameters 
as in Section \ref{sec:noninter_case}.
Furthermore we also consider different ranges of electron-phonon coupling strength: 
from weak to intermediate $\gamma_0/\omega_0 \sim 0.7$ for which the Hartree-Fock 
approximation is valid \cite{Dash:2010,Dash:2011}.

In the following, we study the behaviour of the NE FD relations in 
the presence and in the absence of interaction. We show how
the interaction modifies the FDR of the GFs and of the CC and JJ correlation
functions.

\subsection{Off-resonant transport regime}
\label{sec:interac_offres}

\begin{figure}
  \center
  \includegraphics[clip,width=\mycolumnwidth,height=55mm]{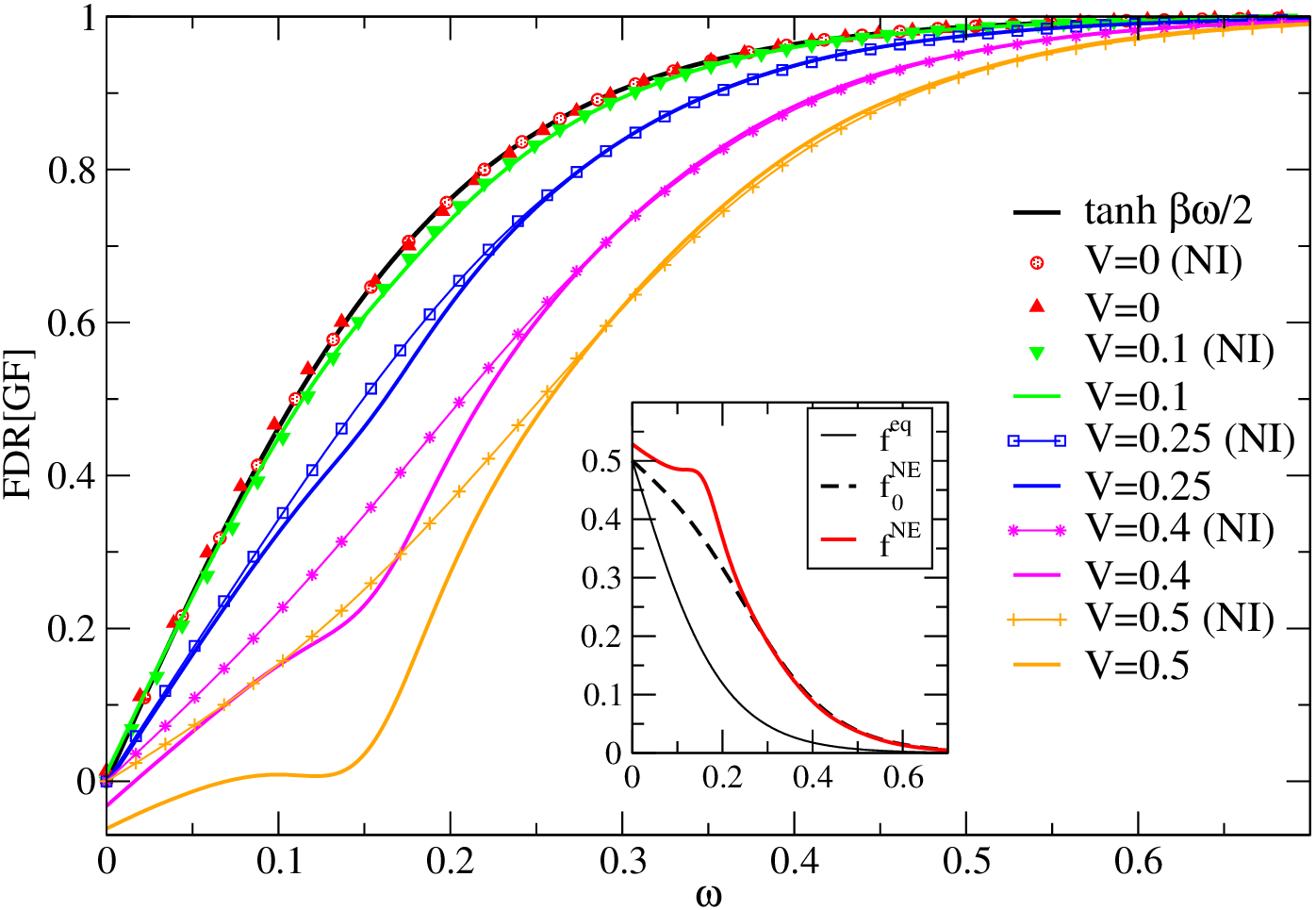}
  \caption{(color online)
FD ratio of the GF, for both the non-interacting (NI), Eq.~(\ref{eq:FDratioG0_withgradT}),
and interacting, Eq.~(\ref{eq:FDratioG}), cases.
Calculations are performed for the off-resonant regime ($\varepsilon_0=0.50$) and different biases $V$. 
At equilibrium, the FD ratio is given by $\tanh \beta\omega/2$ (see $V=0$).
The presence of interaction induces strong deviation from the non-interacting
FD ratio Eq.~(\ref{eq:FDratioG0_withgradT}).
The other parameters are $\gamma_0=0.12, \omega_0=0.3$,  
$t_{0\alpha}=0.15$, $T_\alpha=T_{\rm ph}=0.1$, $\eta_V=0.5$, 
$\varepsilon_\alpha=0, \beta_\alpha=2$.
Inset: Equilibrium ($f^{\rm eq}$) and NE distributions 
($f_0^{\rm NE}$ and $f^{\rm NE}$) for an applied bias $V=0.5>\omega_0$.} 
  \label{fig:5}
\end{figure}

Figure \ref{fig:5} shows the FD ratio of the GF in the off-resonant transport regime, in
the presence and the absence (NI) of interaction. The results for the resonant transport regime 
are given in Appendix \ref{app:interac_res}. 

One can clearly see that the presence of interaction strongly modify the FD ratio, even for
the medium coupling regime ($\gamma_0/\omega_0 =0.4$ in Fig.~\ref{fig:5}).
The deviation from equilibrium are stronger for larger $V$ when
the bias window include a substantial spectral weight of the self-energy $\Sigma_{\rm int}^\lessgtr$.
This is the regime when the single-(quasi)particle representation for quantum transport
breaks down \cite{Ness:2010}.

\begin{figure}
  \center
  \includegraphics[clip,width=\mycolumnwidth,height=55mm]{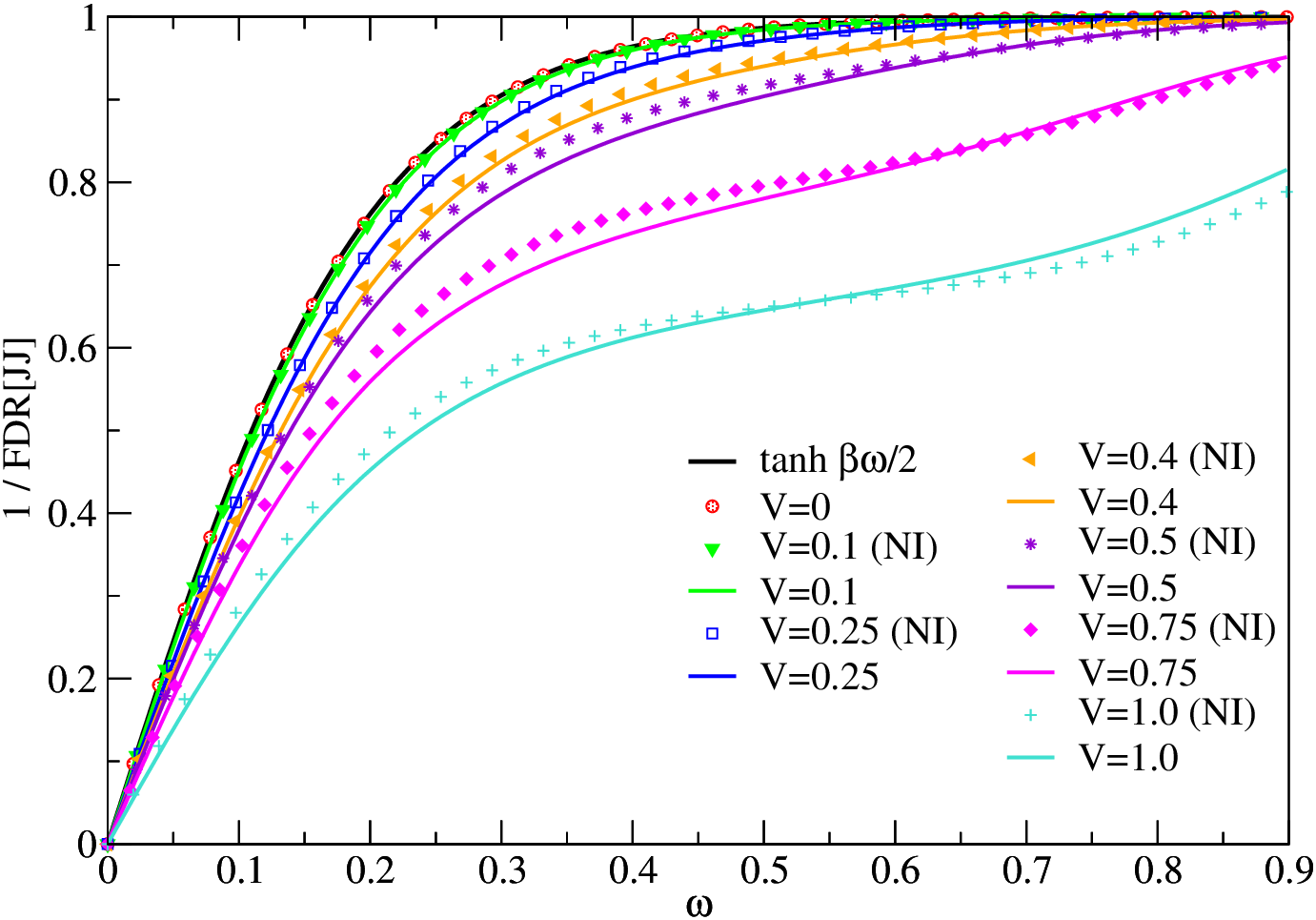} \\
  \includegraphics[clip,width=\mycolumnwidth,height=55mm]{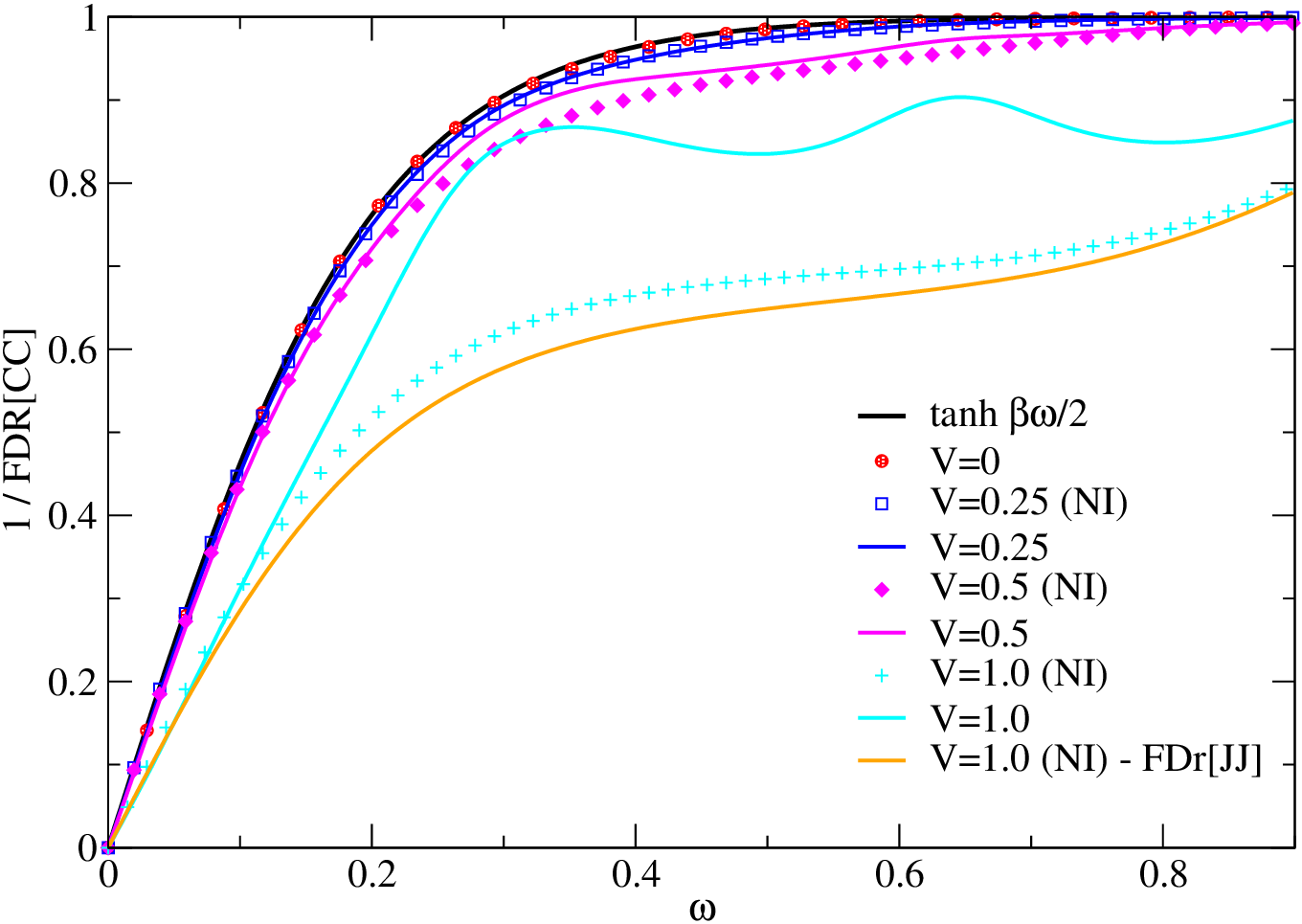}
  \caption{(color online)
(\emph{Top panel}) Inverse FD ratio of the JJ correlation function.
(\emph{Bottom panel}) Inverse FD ratio of the CC correlation function in the off-resonant regime, and
for the non-interacting (NI) and interacting cases.
A curve for FDR[JJ] is also given to clearly show that FDR[CC] $\ne$ FDR[JJ].
The parameters are $\varepsilon_0=0.50$, $\gamma_0=0.12, \omega_0=0.3$,  
$t_{0\alpha}=0.15$, $T_\alpha=T_{\rm ph}=0.1$, $\eta_L=0.5$, 
$\varepsilon_\alpha=0, \beta_\alpha=2$.}
  \label{fig:6}
\end{figure}

At large bias, we can obtain negative values of the FD ratio of the GFs.
This is when the NE MB effects
are not negligible and induce strong modifications of the NE distribution $f^{\rm NE}$
as shown in the inset of Figure \ref{fig:5} for the bias $V=0.5$ larger than the
energy of the phonon $\omega_0=0.3$.

Even for a symmetric coupling to the leads, the MB effects induces a redistribution of the
electron population in $f^{\rm NE}$, as shown by the red solid line in the inset of 
Figure \ref{fig:5}. 
The corresponding NE distribution shows an increase of electron population around
$\omega\sim 0.15$ which correspond to the position of a phonon side-band peak in
the spectral density of the central region.
The NE distribution becomes asymmetric around $\omega=0$ while the distributions 
$f^{\rm eq}$ and $f_0^{\rm NE}$ are still symmetric (as explained
in Section \ref{sec:typicalres_noint}).

These results show that the NE FD relation for the GF is strongly dependent on the 
NE conditions as well as on the MB effects.
It should be noted that however the GF are quantities
not directly accessible to measurements in opposition to the current or the charge.

It is now interesting to see how the FD relations for the 
two-particle correlations are affected by the interaction.
In Figure \ref{fig:6}, we show the inverse FD ratio for the JJ and CC correlation functions for
the off-resonant regime (same parameters as in Fig.~\ref{fig:5}).
The inverse of the FD ratio for two-particle correlation functions is completely different
from the FD ratio of the GFs. Although, in the limit of very small applied bias, it follows 
approximately well the equilibrium behaviour in $\tanh \beta\omega/2$.

However, for finite but small bias, the FDR of the two-particle correlation functions 
with interaction is not well represented by an effective equilibrium form 
in $\tanh \beta_{\rm eff}\omega/2$, as it was the case in the absence of interaction.

More interestingly, the NE FD ratio for the JJ correlations is much less dependent 
on the interactions themselves than the FDR of the GFs. 
More results for the dependence of the FD ratio upon the strength of the interaction
are given in Appendix \ref{app:interac_strength}. They clearly show the weaker 
dependence of FDR[JJ] upon the interaction strength.
The reasons why the FDR for the different correlation functions are different from
each other can be understood from the NE density matrix approach described in 
Section \ref{sec:Yoperator}. 
Indeed, the NE corrections to the equilibrium FDR are obtained from the expansion of
the quantity $ e^{-\beta Y } B e^{\beta Y} $. 
For the GFs, the fermion operator $B$ is $d$ or $d^\dag$, 
for the JJ correlations, the boson-like operator $B$ is $c^\dag_\alpha d$ or 
$d^\dag c_\alpha$ where $c^\dag_\alpha$ ($c_\alpha$) creates (annihilates) an 
electron in the lead $\alpha$, and for the CC relations, the boson-like operator $B$ 
is $d^\dag d$. Hence the series of commutators $[B,Y]$, $[B,[B,Y]]$ ... will be 
different for each different quantity represented by $B$.
 
Therefore, the FDR of the GF and the JJ or CC correlations are not be identical.
Furthermore, for the JJ correlations, one deals with an higher order product 
than for the GFs. 
Hence the expansion of $ e^{-\beta Y } B e^{\beta Y} $ contains higher order powers in 
terms of the interaction coupling parameters (in case our case, in terms of $\gamma_0^2$)
in comparison to the series expansion for the GFs.
For weak to intermediate interaction coupling strengths, one may expect less
effect from the interaction on the FDR of the JJ correlations than on the FD ratio
of the GF.

\subsection{Thermoelectric transport regime}
\label{sec:interac_thermoelec}

\begin{figure}
  \center
  \includegraphics[clip,width=\mycolumnwidth,height=55mm]{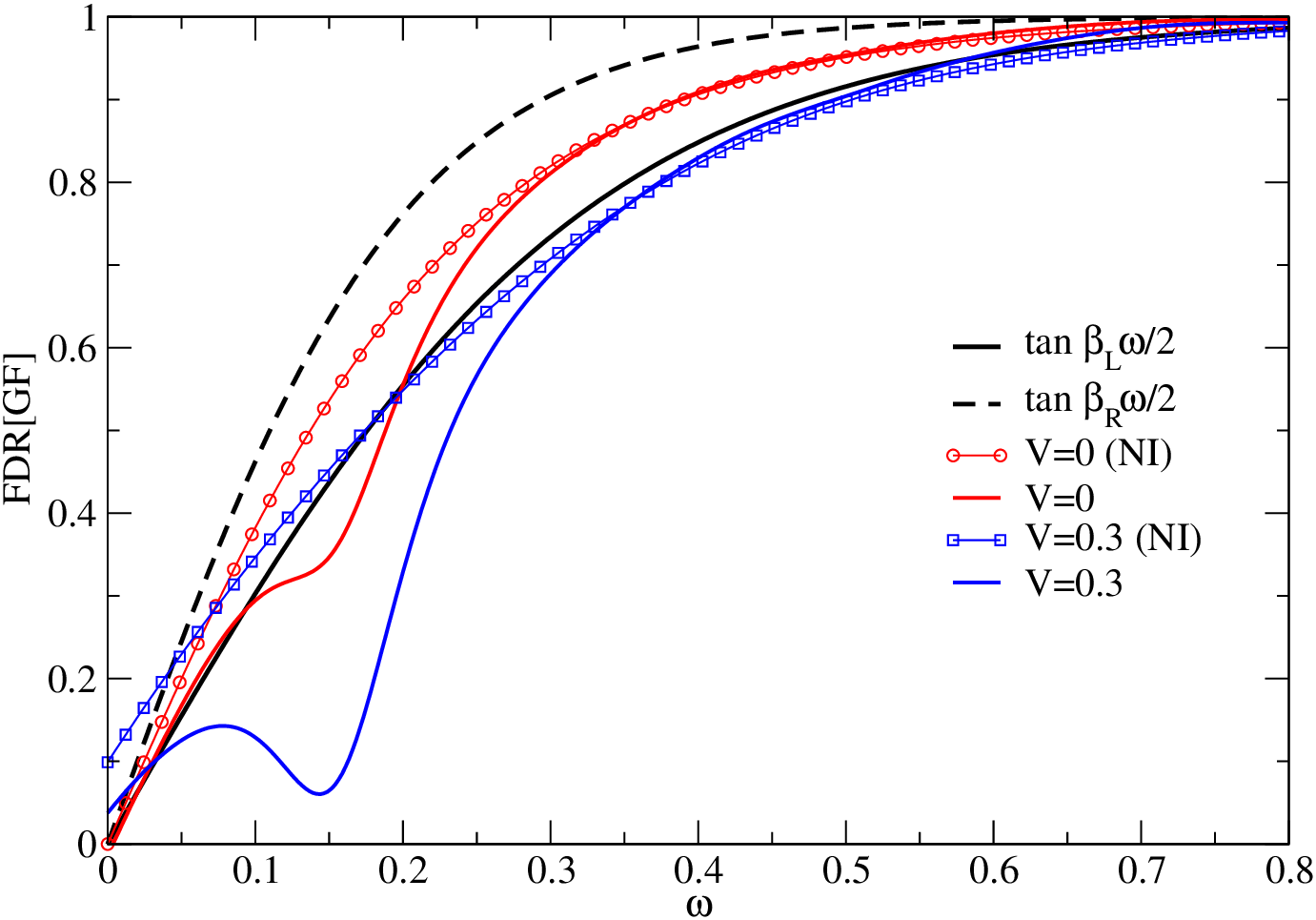} \\
  \includegraphics[clip,width=\mycolumnwidth,height=55mm]{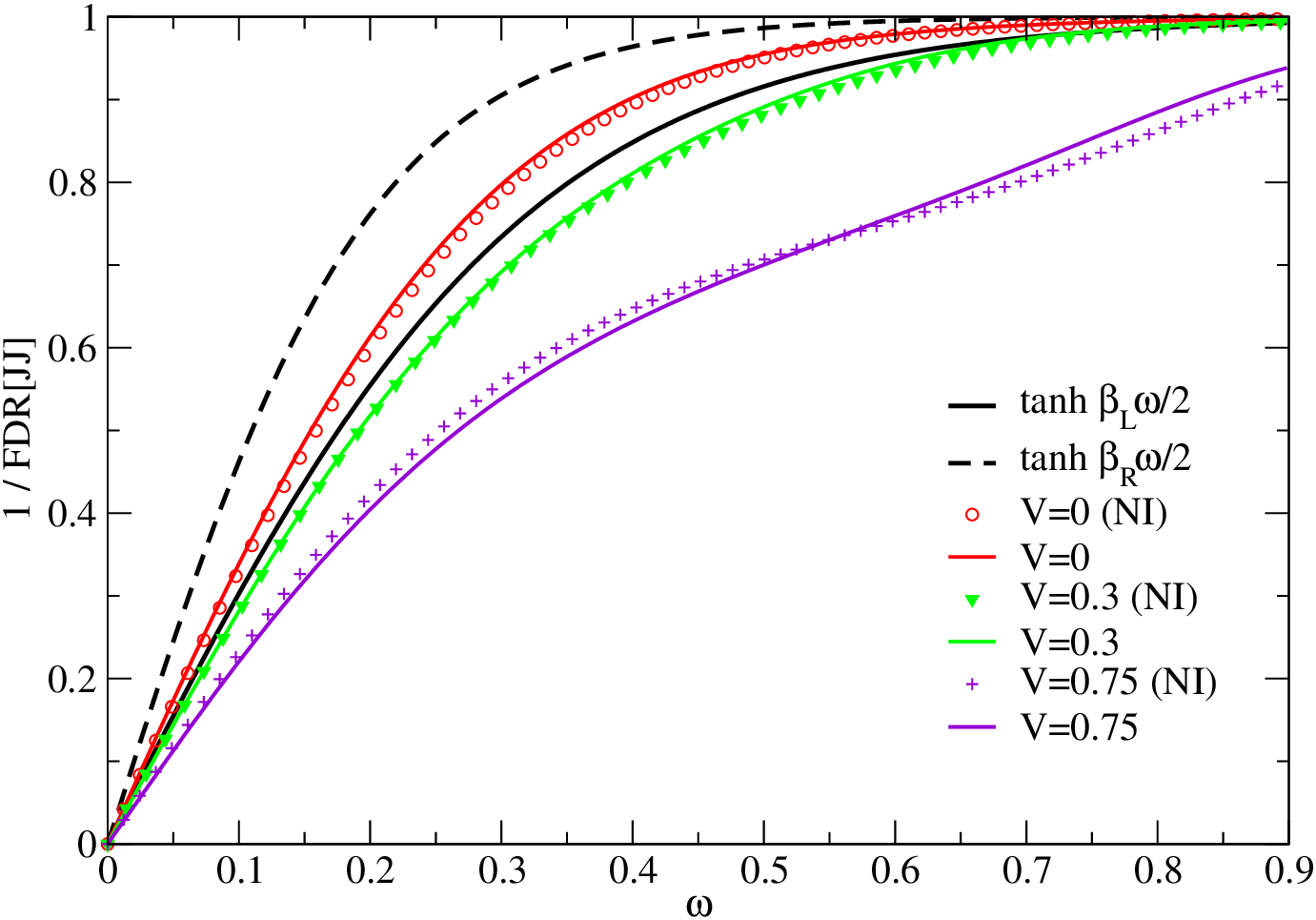} \\
  \includegraphics[clip,width=\mycolumnwidth,height=55mm]{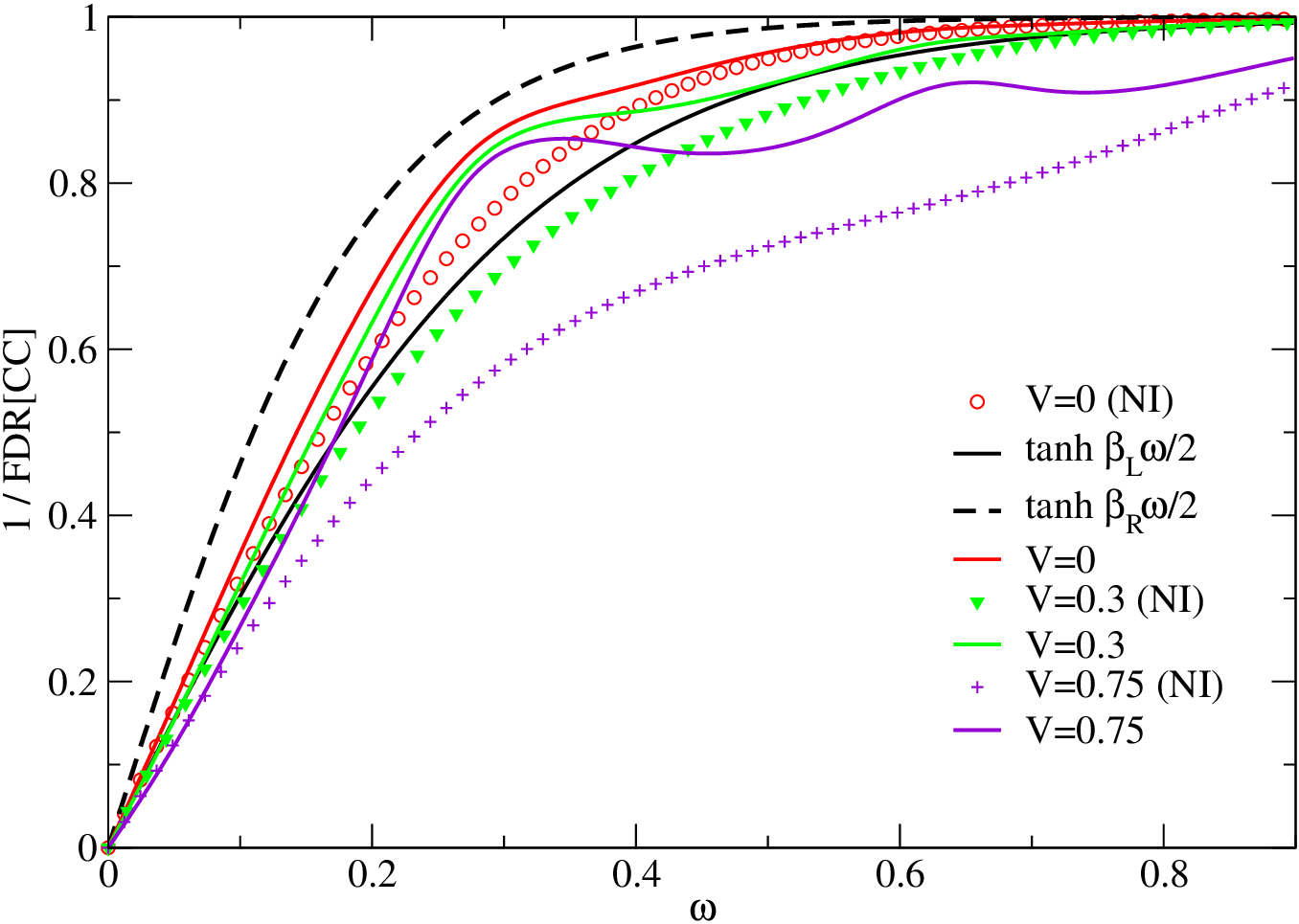}
  \caption{(color online)
FD ratio of the GFs (top panel) and of the JJ correlation functions
(middle panel) and the CC correlation functions (bottom panel) 
in the off-resonant regime ($\varepsilon_0=0.50$) and
for the non-interacting (NI) and interacting cases, and with a temperature gradient 
between the leads ($T_L=0.16$ and $T_R=0.1$).
The other parameters are $\gamma_0=0.12, \omega_0=0.3$,  
$t_{0\alpha}=0.15$, $T_{\rm ph}=0.1$, $\eta_L=0.5$, 
$\varepsilon_\alpha=0, \beta_\alpha=2$.}
  \label{fig:9}
\end{figure}

In this section, we consider the effects of both a chemical potential gradient and
a temperature gradient across the two-terminal junction.
Figure \ref{fig:9} shows the FD ratio of the GFs and the inverse FDR of the
JJ and CC correlation functions for the off-resonant regime (the results for the
resonant regime are given in Appendix \ref{app:interac_thermoelec}).

The FDR of the GFs is once more strongly dependent on both the 
NE and the MB effects, while the FDR of the JJ correlation is 
much less dependent on the interaction.
The presence of the temperature gradient act as if the central region is subjected to
an effective temperature in between $T_L$ and $T_R$. This can be easily understood
from the expression of the FD ratio for the non-interacting GF given by Eq.~(\ref{eq:FDratioG0_withgradT}).
In this equation, $\bar\beta$ (the average of the inverse temperatures $\beta_\alpha$ of 
the leads) plays the role of the $\beta$ factor in the absence of temperature gradient.

Furthermore, as soon as $V \ne 0$, we observe a shift in the FDR of the GFs, i.e. 
FDR[G] $\ne 0$ at $\omega=0$. This shift in the $\omega$-axis can also be
understood from examination of Eq.~(\ref{eq:FDratioG0_withgradT}) and has been explained
in detail in Section \ref{sec:typicalres_noint} for the non-interacting case.

\subsection{Influence of the coupling to the leads}
\label{sec:interac_leadscoupling}

So far, we have considered ``weak'' coupling to the leads in the sense that
the features in the spectral function of the central region, i.e. main 
resonance and phonon side-band peaks, can be resolved 
\cite{Dash:2010,Dash:2011}.
We now consider the limit of strong coupling to the leads where the width of the 
peaks ($\sim \Gamma/2$) in the spectral function is much larger than their 
energy separation ($\sim \omega_0$). The corresponding features in the spectral function 
are almost completely washed-out. 

\begin{figure}
  \center
  \includegraphics[clip,width=\mycolumnwidth,height=55mm]{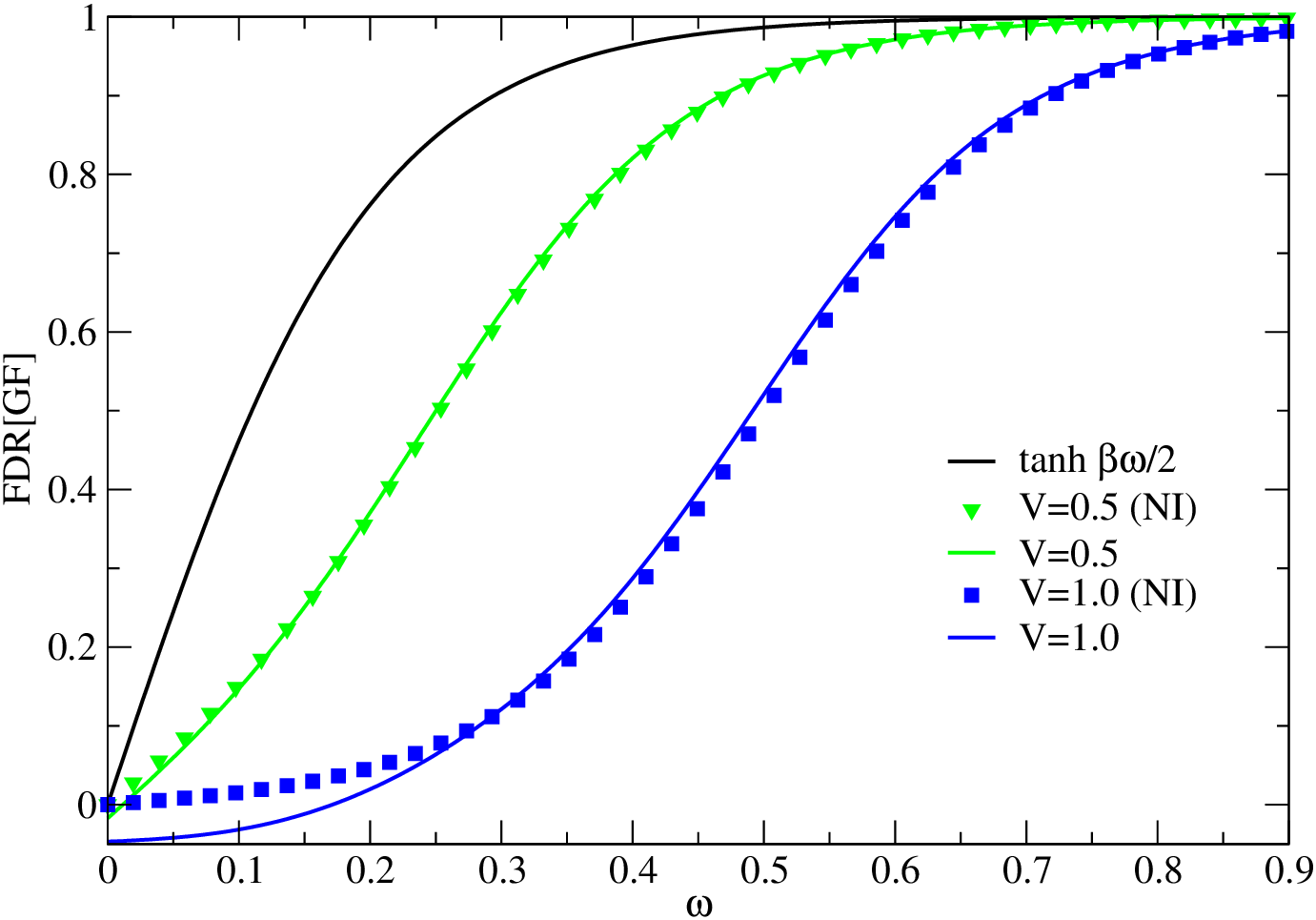} \\
  \includegraphics[clip,width=\mycolumnwidth,height=55mm]{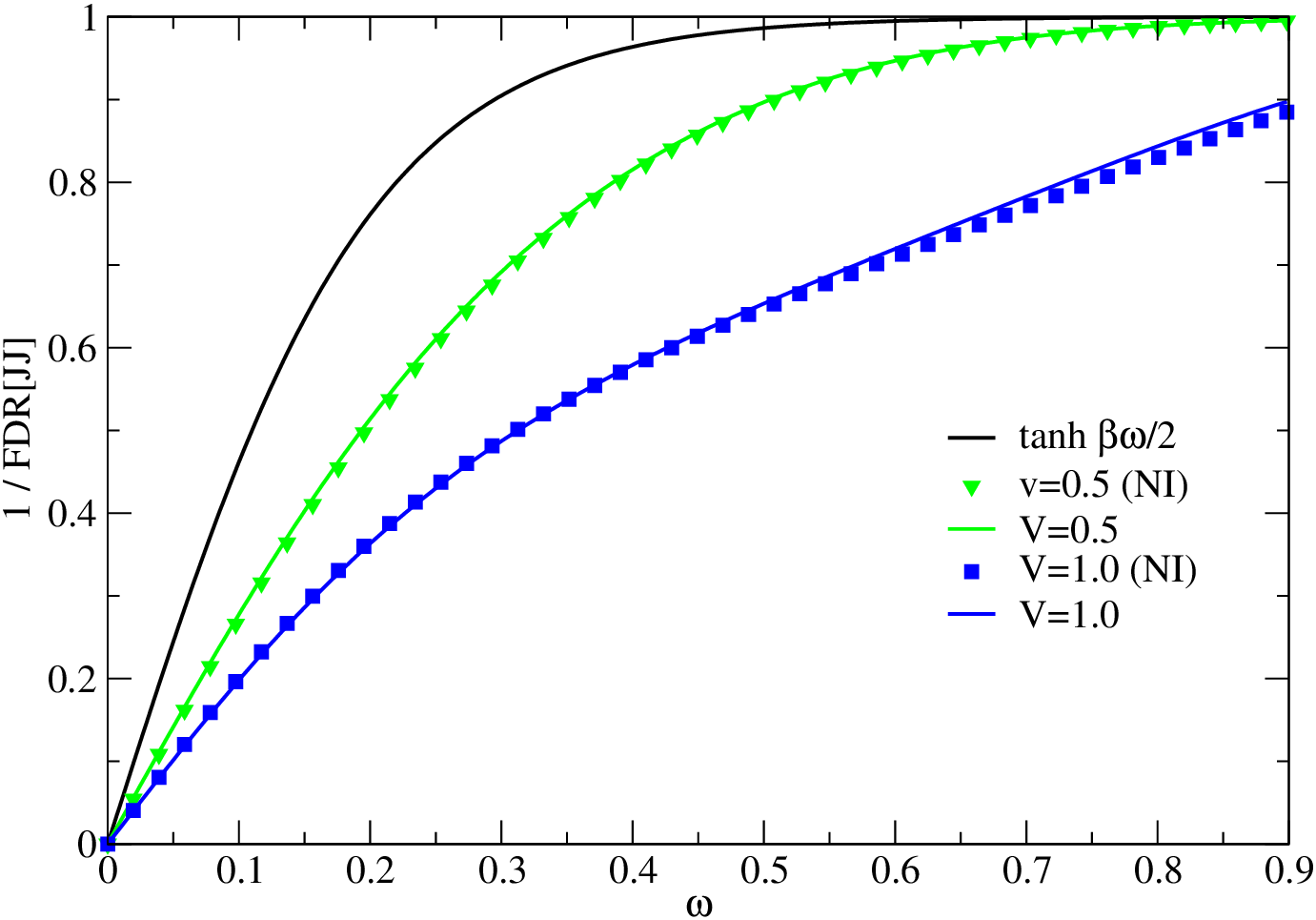} \\
  \includegraphics[clip,width=\mycolumnwidth,height=55mm]{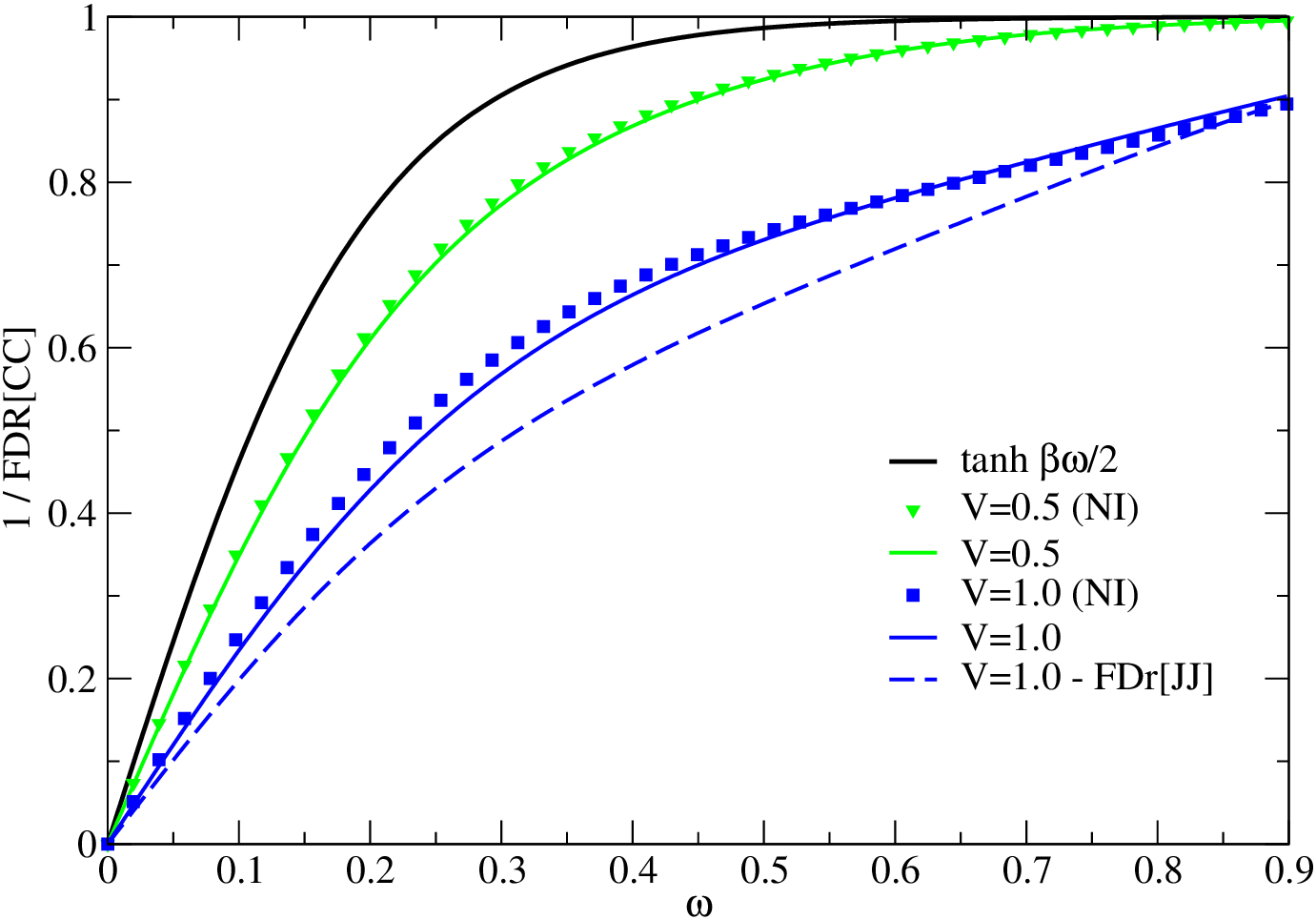}
  \caption{(color online)
FDR of the GFs (top panel) and of the JJ (middle panel)
and CC (bottom panel) correlations. Calculations are performed for 
the off-resonant regime ($\varepsilon_0=5.0$) and with strong coupling
to the leads ($t_{0\alpha}=0.50$).
The coupling is so strong that the spectral features in the NEGFs
are almost completely washed out, and there is almost no difference between
the non-interacting (NI) and interacting results.
The other parameters are $\gamma_0=0.12, \omega_0=0.3$,  
$t_{0\alpha}=0.15$, $T_\alpha=T_{\rm ph}=0.1$, $\eta_L=0.5$, 
$\varepsilon_\alpha=0, \beta_\alpha=2$.}
  \label{fig:10}
\end{figure}

Figure \ref{fig:10} shows the FD ratio of the GFs and the inverse FDR of the
JJ and CC correlation functions in the off-resonant regime, for such a strong coupling
to the leads. 
We can see now that the differences between the interacting and non-interacting results
is much less important and almost non-existent in FDR[JJ] and in FDR[CC].
We can deduce that such differences, and for example the oscillatory-like behaviour
observed in FDR[GF] and 1/FDR[CC] (figs.~\ref{fig:6} and \ref{fig:9}), are related to 
the presence of the peak-like features in the spectral function of the central region.

\section{Discussion}
\label{sec:discuss}

One of the important output of our detailed analysis is that the FD relations for the GFs 
are strongly dependent on both the `forces' ($\Delta\mu$ and/or $\Delta T$) 
driving the system out of equilibrium and on the interaction present in open quantum system. 

However the FD relations for the current-current correlations are much less dependent on 
the interaction itself, at least for the weak to intermediate electron-phonon coupling regime. 
Such a regime corresponds to the
most probable range of coupling strengths for real single-molecule junctions \cite{Dash:2012}. 

The weaker dependence on the interaction of the FDR of the JJ correlations implies that the 
calculated relations for the non-interacting case could be used as master curves for fitting 
experimental results. 
This is important because only the current and the charge, and their fluctuations can
be measured experimentally, and not the GFs themselves.

If the CC and JJ correlation functions are measurable experimentally, as for example by
noise and (linear or not) response measurement, one could use the corresponding FD ratio 
to get information about the local ``microscopic'' properties of the open quantum system.

Indeed, by fitting the experimental FDR onto the master curves (for the non-interacting case), 
one can extract quantitative values of the forces acting effectively on the central region.
Furthermore, other information could be obtained for the strength and symmetry of the coupling 
to the reservoirs (see Sec. \ref{sec:interac_offres} and \ref{sec:interac_leadscoupling}), 
for the strength of the interaction (see Appendix \ref{app:interac_strength}),
for other properties related to the electron-hole symmetry of the system (see Appendix \ref{app:interac_res}).
These are crucial quantities to know in single-molecule nanodevice experiments since they
are not (yet) reproducible in a controlled manner from device to device.

Furthermore, a strong departure from the master curves could also indicate a general breakdown 
of the main hypothesis used in our model: the interactions are present also outside the central
region, or there are more than two energy/particle reservoirs connected to the central 
region.

\section{Conclusion}
\label{sec:ccl}

We have derived FD relations for the one-particle and the two-particle correlation functions 
in the context of quantum transport through a two-terminal device in the steady state regime.
We have also provided numerical applications of our derivations for the case of a single
impurity model in the presence of electron-phonon interaction.
Our calculations are mostly relevant for electron-phonon interacting systems, but are 
not limited only to these processes.
Indeed, they could also be valid for systems with electron-electron interaction when one
considers approximations of the dynamical screening of the interaction, leading to an
effective electron-boson model Hamiltonian \cite{Ness:2011b}.

We have expressed the FD ratio of the GFs (one-particle correlations) for different cases.
We have found a universal form for the FDR in absence of interaction and with a symmetric
coupling to the leads, i.e. the FD ratio depends only on the temperatures of the reservoirs
and on the applied bias. In the case of asymmetry, the FDR depends also on the coupling to
the leads via the $\Gamma_\alpha$ quantity.
In the presence of interactions in the central region, the FDR depends additionally on
these interactions via the self-energies $\Sigma^\lessgtr_{\rm int}$.

The expressions of the FDR for the current-current and charge-charge (two-particle) 
correlations are not obtained in an straightforward analytic manner. We have obtained 
such FD ratios from numerical NE GF calculations.
 
We have shown that the FD ratio for both the one-particle and two-particle correlation
functions are always different from the universal equilibrium FDR.
The FD relations depend on both the `forces' ($\Delta\mu$ and/or $\Delta T$) 
driving the system out of equilibrium and on the interaction. 
However the FD relations for 
the current-current correlations are much less dependent on the interaction itself.
Furthermore, we have shown that, in the linear regime, the FDR of the current-current 
correlations can be described by a effective equilibrium relation, using an effective 
local temperature which is dependent on the applied bias.

We have discussed the possibility of using the FDR calculated for the non-interacting 
case as a master curve for fitting experimental results.
This suggests interesting applications for single-molecule and other nanoscale transport 
experiments where the transport is dominated by a single molecular electronic level.

Measurements of the charge and current, and of their fluctuations can
  provide information about the effective gradients of chemical potential and
  temperature in the central region, and about other properties of the system such as the 
  strength of the coupling to the leads and the strength of the interaction.

\begin{acknowledgments}
The present work has been developed at the UoY.
We gratefully acknowledge L. Arrachea and A.J. Fisher for useful comments, 
and AJF for suggesting the calculation of the charge-charge correlation
functions.
HN thanks Th. Martin for fruitful discussions about the calculations of
the current noise on the Keldysh contour.
\end{acknowledgments}

\appendix

\section{NEGF}
\label{app:sec:NEGF}

The Green's function is defined on the Keldysh time-loop contour $C_K$ as
follows: 
\begin{equation}
\label{eq:defGFonCK}
G_{XY}(t,t')	= - {\rm i} \langle \mathcal{T} c_X(t) c^\dag_Y(t') \rangle,
\end{equation}
where $X,Y$ stands for a composite index for the electronic states in the $L, C$
or $R$ region,  and the time ordering $\mathcal{T}$ of the product
of fermion creation ($c^\dag_Y$) and annihilation ($c_X$) quantum fields
is performed on the time-loop contour \cite{Keldysh:1965,Danielewicz:1984,Chou:1985,vanLeeuwen:2006}.

When the 2 time arguments are on the same branch of $C_K$, we deal with the usual time (anti-time) ordered
Green's function. When the time arguments are on different branches, there is an automatic time ordering knowing
that any time on the backward-time branch (-) is later on the contour $C_K$ than any time on the forward-time
branch (+), i.e.
\begin{equation}
\label{eq:defGFgrtless}
\begin{split}
G_{XY}^{-+} \equiv G_{XY}^>(t,t')	& = - {\rm i} \langle c_X(t) c^\dag_Y(t') \rangle, \\
G_{XY}^{+-} \equiv G_{XY}^<(t,t')	& =   {\rm i} \langle c^\dag_Y(t') c_X(t) \rangle,
\end{split}
\end{equation}

\section{KMS and FDR for the non-interacting case}
\label{app:FDRnoint}

In this section, we derive the expressions for the KMS and FD ratios for an open quantum system at 
a unique temperature $T$ and in the presence of an applied bias $V$. Some of the results can be
found in similar or different forms in the literature \cite{Kirchner:2009,Blanter:2000}.

For symmetric coupling to the leads, $\Gamma_L=\Gamma_R$ and $\mu_{L,R}=\mu^{\rm eq}\pm V/2$, we obtain
the following results for the FD and KMS ratio after algebraic manipulations:
\begin{equation}
\label{eq:FDratioG0sym}
{\rm FDR}[G_0]=\frac{\sinh \beta\bar\omega}{\cosh \beta\bar\omega + \cosh \beta V/2} \ ,
\end{equation}
and
\begin{equation}
\label{eq:KMSratioG0sym}
\frac{G_0^<}{G_0^>}=-\frac{e^{-\beta\bar\omega}+\cosh \beta V/2}{e^{\beta\bar\omega} + \cosh \beta V/2} \ .
\end{equation}
Interestingly, in the symmetric coupling case, both the KMS and FD ratios are independent 
of the coupling to the leads $\Gamma_\alpha$.

At equilibrium, when $V=0$, we recover the expected results: 
\begin{equation}
\frac{G_0^<}{G_0^>}=-\frac{e^{-\beta\bar\omega}+1}{e^{\beta\bar\omega} + 1}=-e^{-\beta\bar\omega} \nonumber
\end{equation}
and 
\begin{equation}
\begin{split}
{\rm FDR}[G_0] & =
\frac{ \sinh \beta\bar\omega }{ \cosh \beta\bar\omega + 1 } \\
& =\frac{(e^{\beta\bar\omega/2}+e^{-\beta\bar\omega/2})(e^{\beta\bar\omega/2}-e^{-\beta\bar\omega/2})}
{(e^{\beta\bar\omega/2}+e^{-\beta\bar\omega/2})^2} \\
& =\tanh \left( \beta\bar\omega/2 \right)  . \nonumber
\end{split}
\end{equation}

In general, the left and right contacts are different ($\Gamma_L \ne \Gamma_R$) 
and there are asymmetric potential drops at the contacts, i.e.  
$\mu_\alpha=\mu^{\rm eq}+\eta_\alpha V$, with the condition $\Delta\mu=\mu_L-\mu_R=V$ ($\eta_L-\eta_R=1$). 
In such conditions, we find that the corresponding FD ratio Eq.~(\ref{eq:FDratioG0}) is given
explicitly by:
\begin{equation}
\label{eq:FDratioG0gen}
{\rm FDR}[G_0]=\frac{\sinh \beta(\bar\omega-\bar\eta V) - \bar\Gamma_{L-R} \sinh \beta V/2}
                        {\cosh \beta(\bar\omega-\bar\eta V) + \cosh \beta V/2} \ ,
\end{equation}
with $\bar\eta=(\eta_L+\eta_R)/2$ and $\bar\Gamma_{L-R}=\bar\Gamma_L - \bar\Gamma_R$ with 
$\bar\Gamma_\alpha=\Gamma_\alpha/(\Gamma_L+\Gamma_R)$. 
For the KMS ratio we obtain the following expression:
\begin{equation}
\label{eq:KMSratioG0gen}
\frac{G^<}{G^>}=-
\frac{e^{-\beta(\bar\omega-\bar\eta V)} + \cosh \beta V/2 + \bar\Gamma_{L-R} \sinh \beta V/2}
     {e^{ \beta(\bar\omega-\bar\eta V)} + \cosh \beta V/2 - \bar\Gamma_{L-R} \sinh \beta V/2} \ .
\end{equation}

Therefore, in the general case, both the FD and KMS ratios are dependent on the coupling 
to the leads via the imaginary part of the leads' self-energies $\Gamma_\alpha(\omega)$.

\section{The two-particle correlation functions}
\label{app:2Pcorrel}

In order to make full use of the Keldysh formalism, we rewrite the
correlation functions with an appropriate time-ordering on the time-loop contour \cite{ThM:2001,Kamenev:2011}, 
i.e.
\begin{equation}
\label{eq:SonCK}
\tilde S^X_{\alpha\beta}(t_1,t_2) = \sum_{\zeta=\pm} \langle \mathcal{T} X_\alpha(t_1^\zeta) X_\beta(t_2^{-\zeta}) \rangle \ ,
\end{equation}
where $\zeta$ indicates the location on the the backward-time branch $(-)$ or forward-time branch $(+)$
of the time-loop contour $C_K$.
Knowing that any time of the branch $(-)$ is always later than any time on the branch $(+)$, we get 
the proper ordering
\begin{equation}
\tilde S^X_{\alpha\beta}(t,t') = \langle X_\alpha(t) X_\beta(t') + X_\beta(t') X_\alpha(t)\rangle \ ,
\end{equation}
with $t$ on branch (-) and $t'$ on branch (+) in the first correlator 
$\langle X_\alpha(t) X_\beta(t')\rangle $, and $t'$ on branch (-) and $t$ on branch (+) in the second 
correlator $\langle X_\beta(t') X_\alpha(t)\rangle $.

With such rules, we redefine the two-particle correlation function $S^X_{\alpha\beta}(t,t')$ and
response function $R^X_{\alpha\beta}(t,t')$ (with a negative sign between the two correlators).
In their evaluation, we have to calculate
averages of strings of product of four fermion operators such as 
$\langle \mathcal{T} c^\dag_X(t) c_Y(t) c^\dag_W(t') c_Z(t')\rangle$ with
$X,Y,W,Z$ indices representing one of the three $L,C,R$ region.

We use the Wick's theorem on the time-loop contour $C_K$ with $t$ on branch $(-)$ 
and $t'$ on branch $(+)$ and we can decompose the products of operators to obtain:
\begin{equation}
\label{eq:Wick_1}
\begin{split}
& \langle \mathcal{T} c^\dag_X(t) c_Y(t) c^\dag_W(t') c_Z(t')\rangle \\
& =  \langle \mathcal{T} c^\dag_X(t) c_Y(t) \rangle \ \langle \mathcal{T} c^\dag_W(t') c_Z(t')\rangle \\
& +  \langle \mathcal{T} c^\dag_X(t) c_Z(t') \rangle \ \langle \mathcal{T} c_Y(t) c^\dag_W(t')\rangle \ .
\end{split}
\end{equation}
Using the definition of the NEGF and the compact notation $A(t)=c^\dag_X(t) c_Y(t)$ and  
$B(t')=c^\dag_W(t') c_Z(t')$, we find that
\begin{equation}
\label{eq:Wick_2}
\langle A(t) B(t')\rangle =
\langle A(t) \rangle \ \langle B(t')\rangle + G_{YW}^>(t,t') \ G_{ZX}^<(t',t) ,
\end{equation}
where an implicit time-ordering on $C_K$ is used in the averages. 
An equivalent relation for $\langle B(t') A(t) \rangle$ is obtained with, this time,  
$t'$ on branch (-) and $t$ on branch (+) :
\begin{equation}
\label{eq:Wick_3}
\langle B(t') A(t) \rangle =
\langle B(t') \rangle \ \langle A(t) \rangle + G_{ZX}^>(t',t) \ G_{YW}^<(t,t') .
\end{equation}
 
The first term in the right hand side of Eq.~(\ref{eq:Wick_2}) and Eq.~(\ref{eq:Wick_3}) 
corresponds to the product of the averaged quantities $A$ and $B$. 
It disappears in the calculation of the commutator 
$\langle [A(t),B(t')]\rangle$ and of the anticommutator  $\langle \{ \delta A(t),\delta B(t')\}\rangle$
since $\delta X(t) = X(t) - \langle X(t) \rangle$.
Hence, we are left with the evaluation of the crossed terms $G_{YW}^>(t,t') \ G_{ZX}^<(t',t)$ 
and $G_{ZX}^>(t',t) \ G_{YW}^<(t,t')$ in the calculation of the two-particle correlation and response functions.

The Green's functions $G_{XY}^\gtrless(t_1,t_2)$ are obtained from the corresponding Dyson equation
on $C_K$
\begin{equation}
\label{eq:GFgtrless}
\begin{split}
 G_{XY}^\gtrless(t_1,t_2) & = g_{XX}^\gtrless(t_1,t_2)\delta_{XY} + (g \Sigma G)_{XY}^\gtrless(t_1,t_2) \\
                          & = g_{XX}^\gtrless(t_1,t_2)\delta_{XY} + (G \Sigma g)_{XY}^\gtrless(t_1,t_2) .
\end{split}
\end{equation}

We evaluate all the necessary matrix elements for the GFs and self-energies using
the Langreth rules of decomposition on the time-loop contour \cite{vanLeeuwen:2006}.
We keep in mind that the off-diagonal elements of $\Sigma_{XY}$ are given by the hopping matrix elements
between the central region $C$ and the $L$ or $R$ lead. 

We finally obtain expressions in terms of quantities (GFs and self-energies) defined only in the central 
region $C$. 
After Fourier transform of the different products $G_{YW}^>(t-t')  G_{ZX}^<(t'-t)$, we end up with traces 
over the states in the region $C$ as given by Eqs.~(\ref{eq:jjab_w}) and (\ref{eq:SRtot_jj_w}) for the current, 
and Eq.~(\ref{eq:NNCC_w}) for the charge respectively.

\begin{figure}
  \center
  \includegraphics[clip,width=\mycolumnwidth,height=55mm]{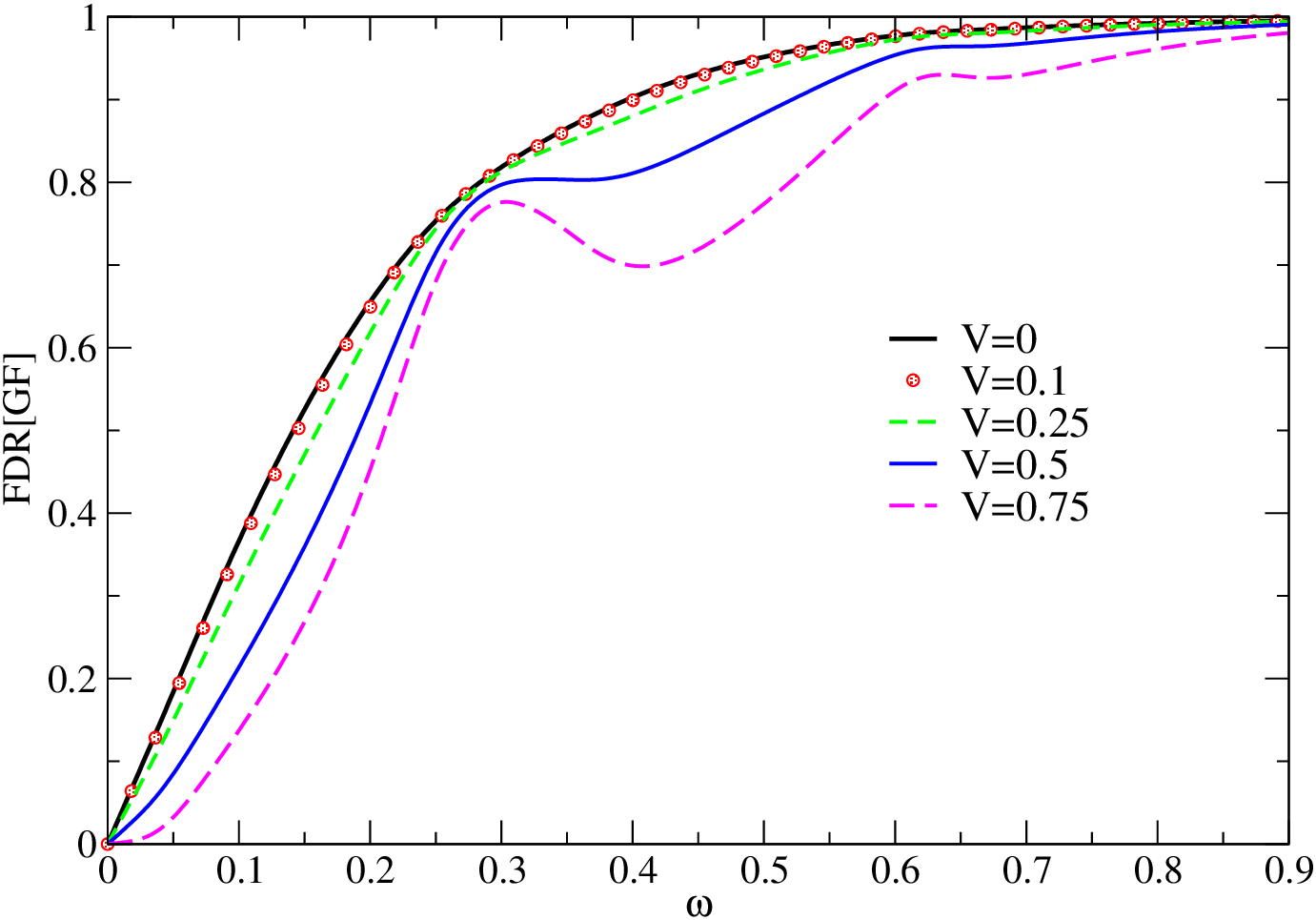} \\
  \includegraphics[clip,width=\mycolumnwidth,height=55mm]{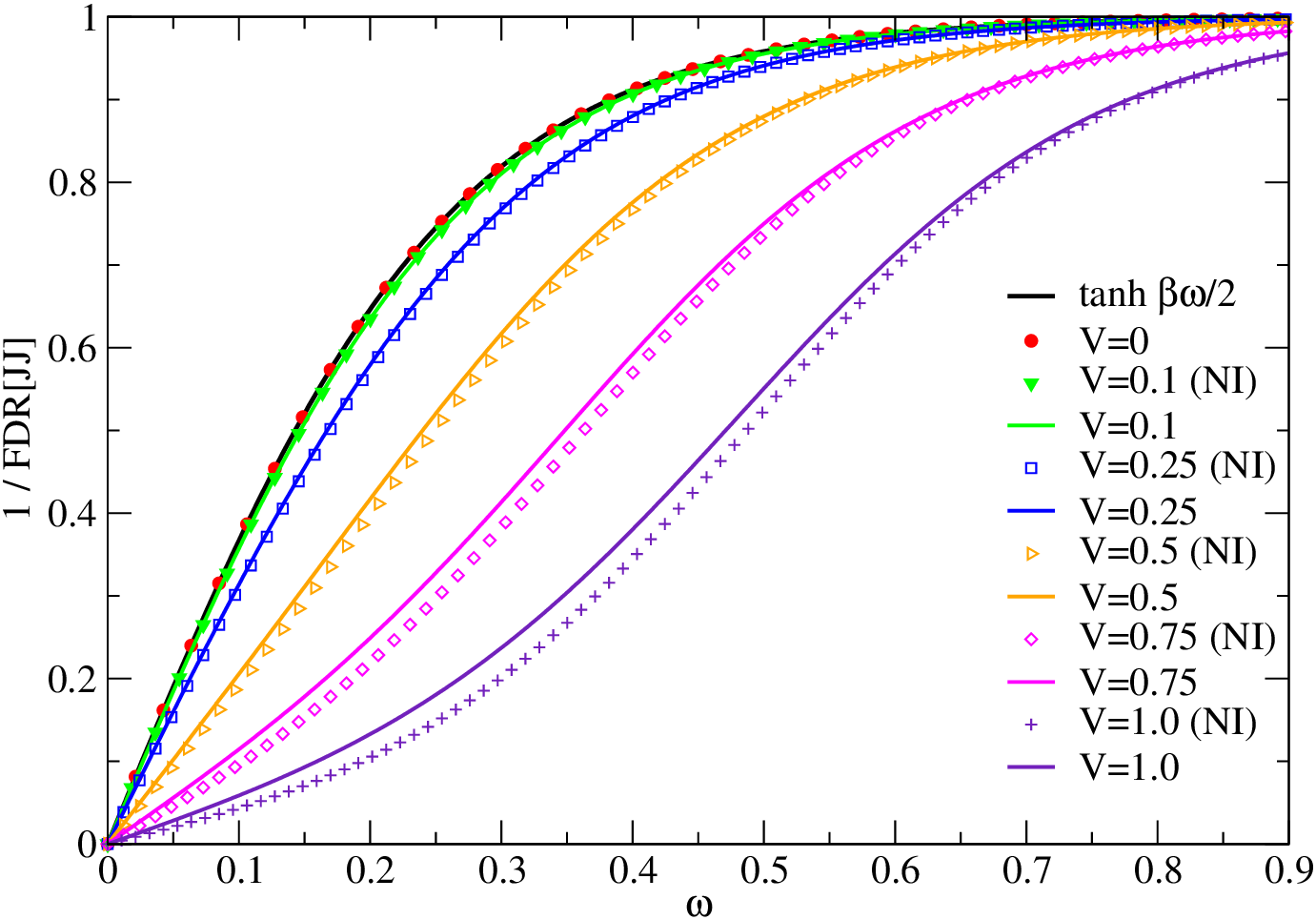} \\
  \includegraphics[clip,width=\mycolumnwidth,height=55mm]{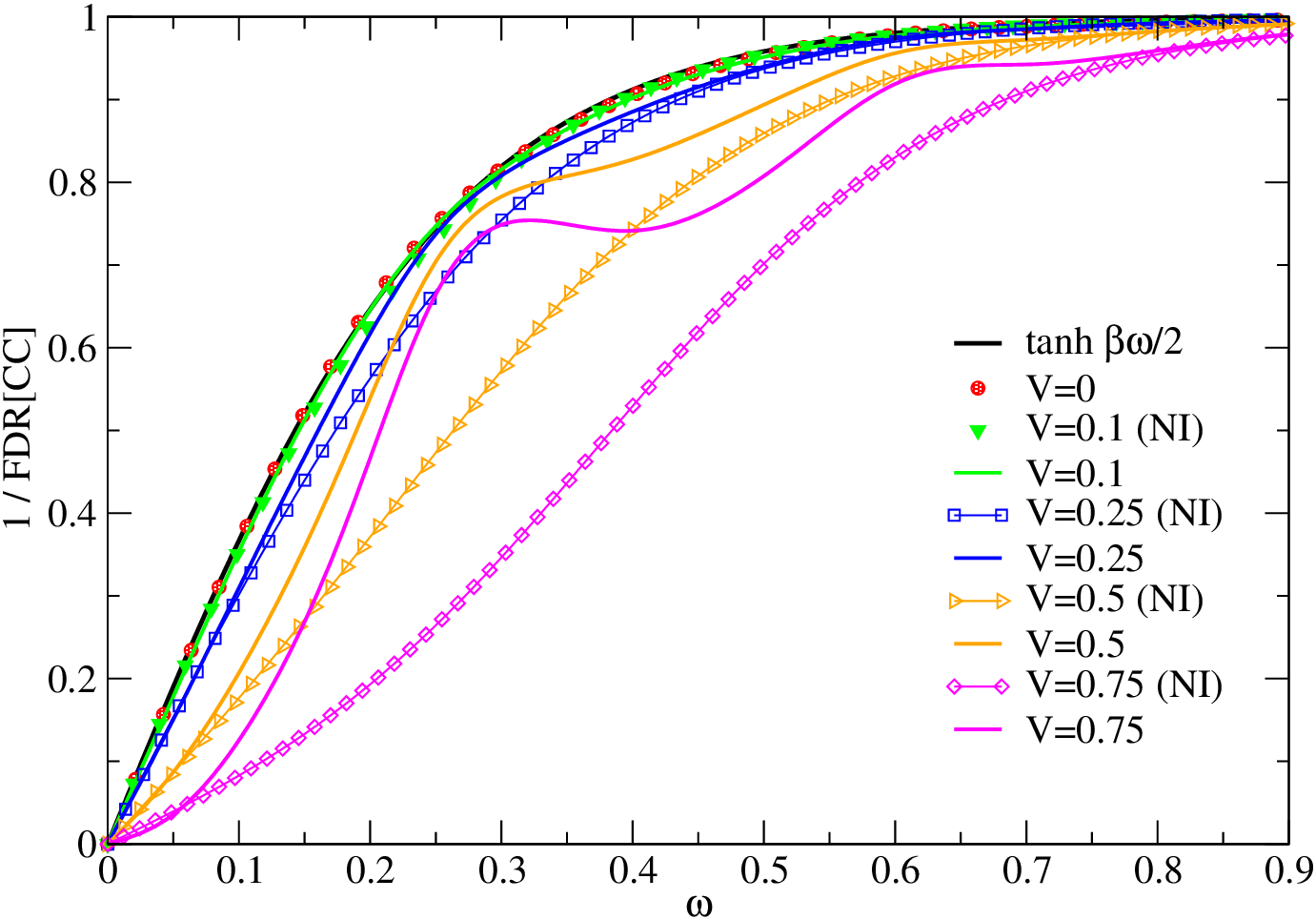}
  \caption{(color online)
FDR of the GFs (top panel) and the inverse FDR of the the JJ (middle panel)
and CC (bottom panel) correlations functions 
in the resonant regime ($\varepsilon_0=0.0$),
in the presence and the absence (NI) of interaction.
The calculations are performed without the Hartree self-energy, so that the system is fully electron-hole 
symmetric for symmetric coupling to the leads.
The other parameters are $\gamma_0=0.12, \omega_0=0.3$,  
$t_{0\alpha}=0.15$, $T_\alpha=T_{\rm ph}=0.13$, $\eta_L=0.5$, 
$\varepsilon_\alpha=0, \beta_\alpha=2$.}
  \label{fig:7}
\end{figure}

\section{The current-current correlation functions}
\label{app:2Pcorrel_JJ}

The full expression for the current-current correlation functions Eq.~(\ref{SRab_w}) is given by
(see Refs.~[\onlinecite{LundBo:1996,Ding:1997,Zhu:2003,Galperin:2007}]):
\begin{equation}
\label{eq:jjab_w}
\begin{split}
\langle j_\alpha j_\beta \rangle _\omega^-
= & \frac{e^2}{\hbar^2}
\int \frac{{\rm d}u}{2\pi}\ {\rm Tr}_C \left[  \right. \\
&\ G^>(u)\ \left(  \Sigma_\alpha\ \delta_{\alpha\beta} + \Sigma_\beta G \Sigma_\alpha \right)^<(u-\omega) \\
+ & \left( \Sigma_\alpha\ \delta_{\alpha\beta} + \Sigma_\alpha G \Sigma_\beta \right)^>(u)\ G^<(u-\omega)  \\
- & \left( G \Sigma_\beta \right)^>(u)\ \left(G \Sigma_\alpha \right)^<(u-\omega) \\ 
- & \left. \left( \Sigma_\alpha G\right)^>(u)\ \left(\Sigma_\beta G \right)^<(u-\omega)\ \right] \ .
\end{split}
\end{equation}
The expression for $\langle j_\beta j_\alpha \rangle _\omega^+$ is obtained from 
$\langle j_\alpha j_\beta \rangle _\omega^-$ by swapping the index $\alpha\leftrightarrow\beta$ 
and taking the energy argument $(u+\omega)$ instead of $(u-\omega)$  
The two-particle correlation functions are bosonic by nature, hence the JJ correlation and response 
functions follows, at equilibrium, the FD relation given by Eq.~(\ref{eq:equiFDT_boson}).

For practical applications, we also consider the total noise 
$S^J(\omega)$ and response function $R^J(\omega)$ obtained from the symmetrized
current $J=(J_L-J_R)/2$. The different quantities $X=S^J, R^J$ are then obtained from
the following linear combination
$X(\omega)=(X_{LL}-X_{LR}-X_{RL}+X_{RR})/4$.
We obtain the corresponding symmetrized version 
$\langle j j \rangle_\omega^{\mp} = 
 \langle j_L j_L \rangle _\omega^\mp
-\langle j_L j_R \rangle _\omega^\mp
-\langle j_R j_L \rangle _\omega^\mp
+\langle j_R j_R \rangle _\omega^\mp$ which is given by
\begin{equation}
\label{eq:SRtot_jj_w}
\begin{split}
\langle j j \rangle _\omega^{\mp}
= & \frac{1}{4} \frac{e^2}{\hbar^2} \int \frac{{\rm d}u}{2\pi} {\rm Tr}_C \left[ \right. \\
& \  G^>(u) \left(\Sigma_{L + R} + \Sigma_{L - R} G \Sigma_{L - R} \right)^<(u\mp\omega) \\
+ &         \left(\Sigma_{L + R} + \Sigma_{L - R} G \Sigma_{L - R} \right)^>(u)\ G^<(u\mp\omega)   \\ 
- &         \left(G \Sigma_{L - R} \right)^>(u)\ \left(G \Sigma_{L - R} \right)^<(u\mp\omega) \\
- & \left.  \left(\Sigma_{L - R} G\right)^>(u)\ \left(\Sigma_{L - R} G \right)^<(u\mp\omega)\ \right] ,
\end{split}
\end{equation}
where $\Sigma_{\alpha\pm\beta}=\Sigma_\alpha\pm\Sigma_\beta$.

\section{Resonant transport regime}
\label{app:interac_res}

In Figure \ref{fig:7}, we show the FD ratio for the GFs and the inverse the FD ratio for JJ and CC 
correlation functions in the resonant regime with full electron-hole symmetry (i.e. no Hartree 
self-energy in the calculations [\onlinecite{Dash:2010}]). 
In this case, the spectral function of the central region always presents an electron-hole 
symmetry.
Furthermore, we can see that FDR[GF] is always positive as the NE distribution $f^{\rm NE}$ presents
the same electron-hole symmetry as the non-interacting and equilibrium distribution
$f_0^{\rm NE}$ and $f^{\rm eq}$.

We observe that, similarly to the off-resonant case, the inverse FDR of the JJ correlation 
functions does not have the same behaviour as FDR[GF], and 
that FDR[JJ] is also different from FDR[CC].
The NE FD ratio for the JJ correlations is still much less dependent on the interaction 
than FDR[GF], at least in the medium coupling strength of the interaction.
Furthermore the FDR[JJ] seems even less dependent on the (electron-hole symmetric) interaction
than it is for the off-resonant case (except for large biases $V \ge 1.0$).

\begin{figure}
  \center
  \includegraphics[clip,width=\mycolumnwidth,height=55mm]{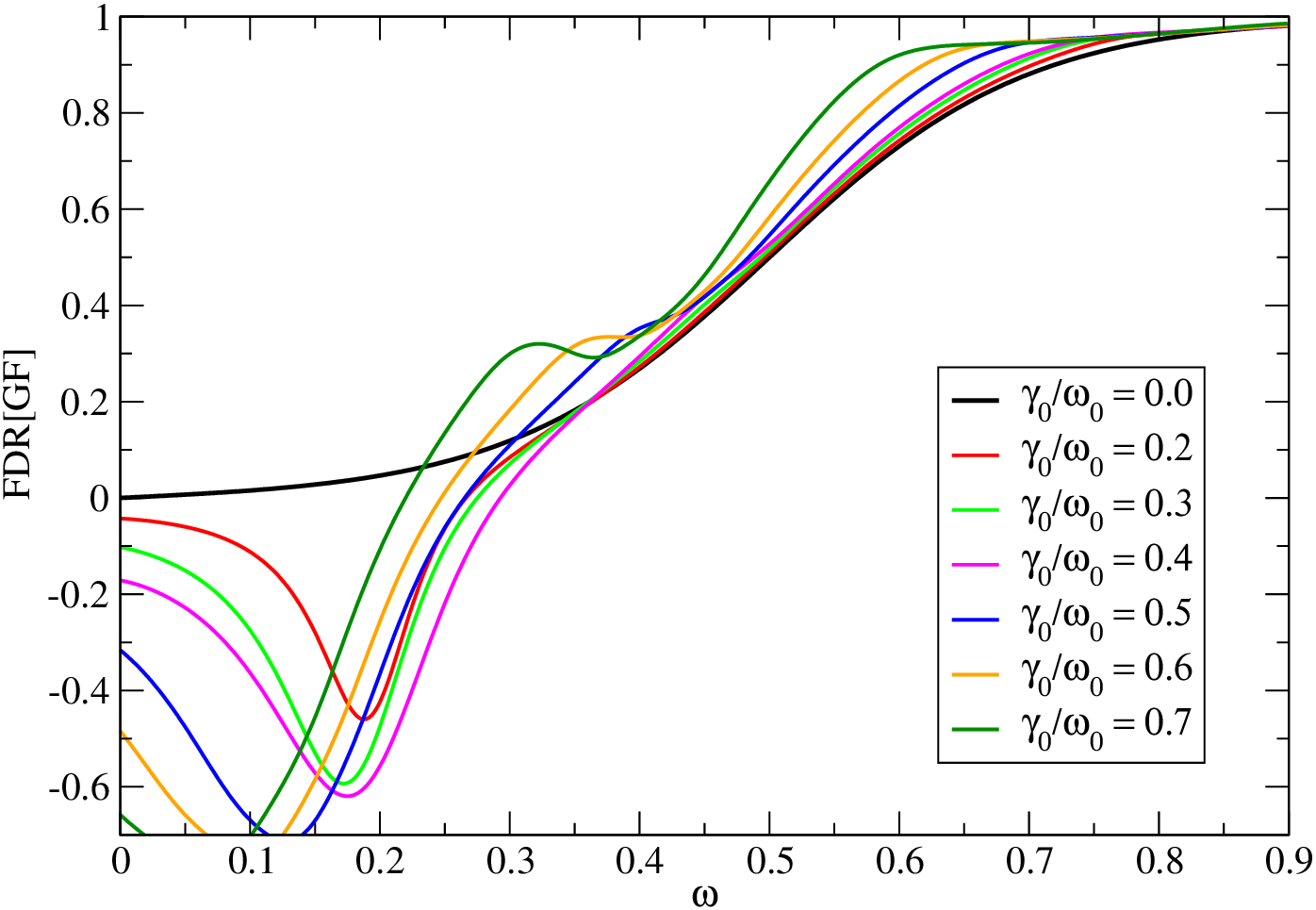} \\
  \includegraphics[clip,width=\mycolumnwidth,height=55mm]{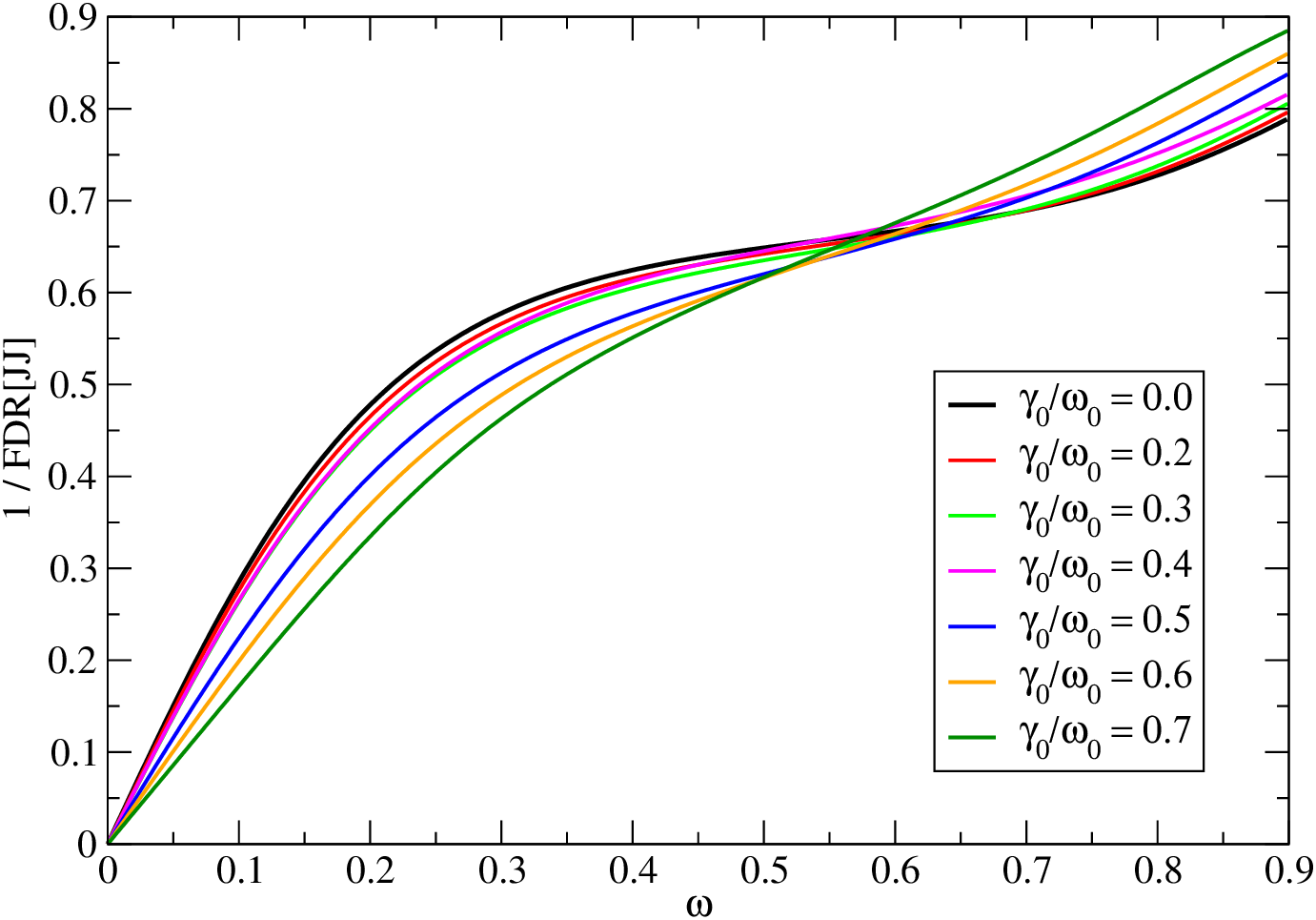} \\
  \includegraphics[clip,width=\mycolumnwidth,height=55mm]{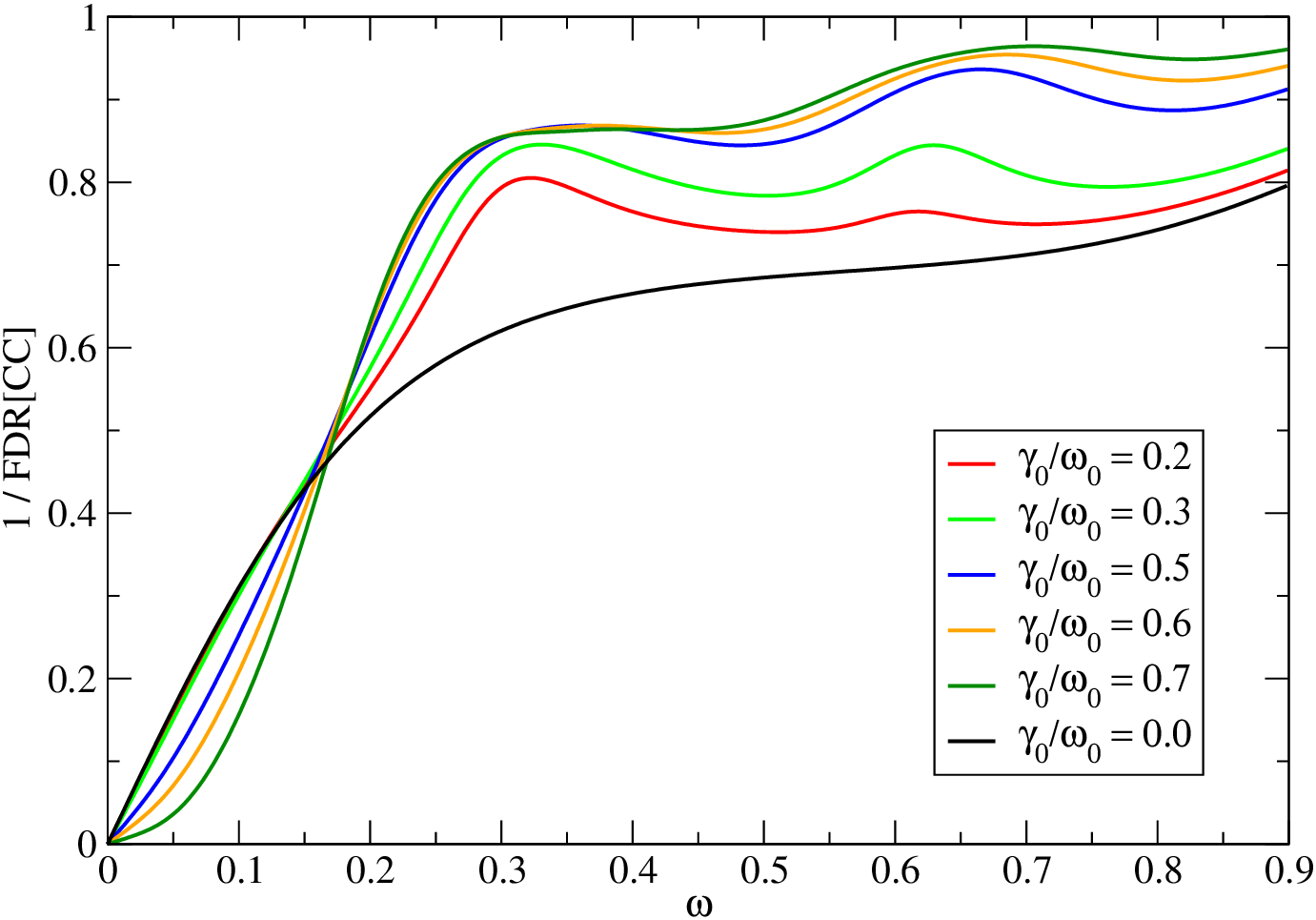}
  \caption{(color online)
FDR of the GFs (top panel), JJ correlations (middle panel)
and CC correlations (bottom panel) 
in the off-resonant regime ($\varepsilon_0=5.0$) for a applied bias $V=1.0$.
Different values of the electron-phonon coupling strength $\gamma_0$
(with $\omega_0=0.3$) are shown.
The other parameters are   
$t_{0\alpha}=0.15$, $T_\alpha=T_{\rm ph}=0.1$, $\eta_L=0.5$,
$\varepsilon_\alpha=0, \beta_\alpha=2$.}
  \label{fig:11}
\end{figure}

\section{Influence of the interaction strength}
\label{app:interac_strength}

In this section, we analyse the dependence of the FD ratios upon the strength of
the interaction.
The results of the calculations for different electron-phonon coupling strength $\gamma_0$
are shown in Figure \ref{fig:11}.
The dependence of the FDR upon the strength of the interaction is clearly shown
in Figure \ref{fig:11}.
The FDR of the GFs is still much more dependent on the interaction strength
that the FDR of the JJ correlations.

For the current-current correlations functions, one can see that FDR deviates from the
non-interacting case only for large electron-phonon coupling ($\gamma_0/\omega_0 > 0.5$).
In such cases, the validity of the lowest order expansion for the interaction self-energies becomes
questionable (see Refs.~[\onlinecite{Dash:2010,Dash:2011}]).

The dependence of FDR[CC] on the interaction is more important than for FDR[JJ]. 
This is probably due to the fact that, in our model, the electron-phonon 
interaction is mediated via the charge density (the local operators $d^\dag d$) in the central 
region. Hence the dependence on the interaction is stronger in the CC correlations than in the 
JJ correlations, since the current is a non-local operator,
not directly related to the charge density in the central region.

\begin{figure}
  \center
  \includegraphics[clip,width=\mycolumnwidth,height=55mm]{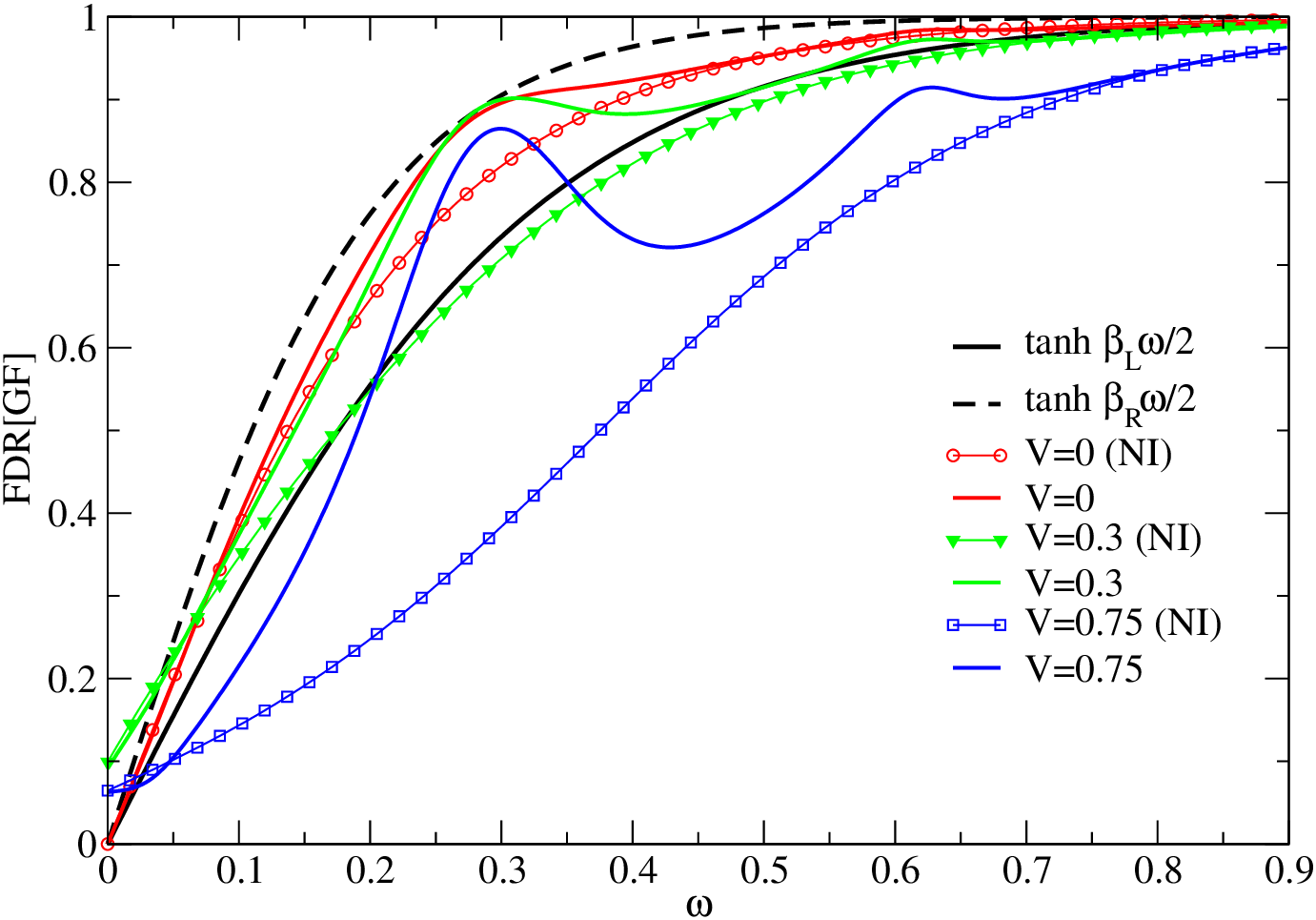} \\
  \includegraphics[clip,width=\mycolumnwidth,height=55mm]{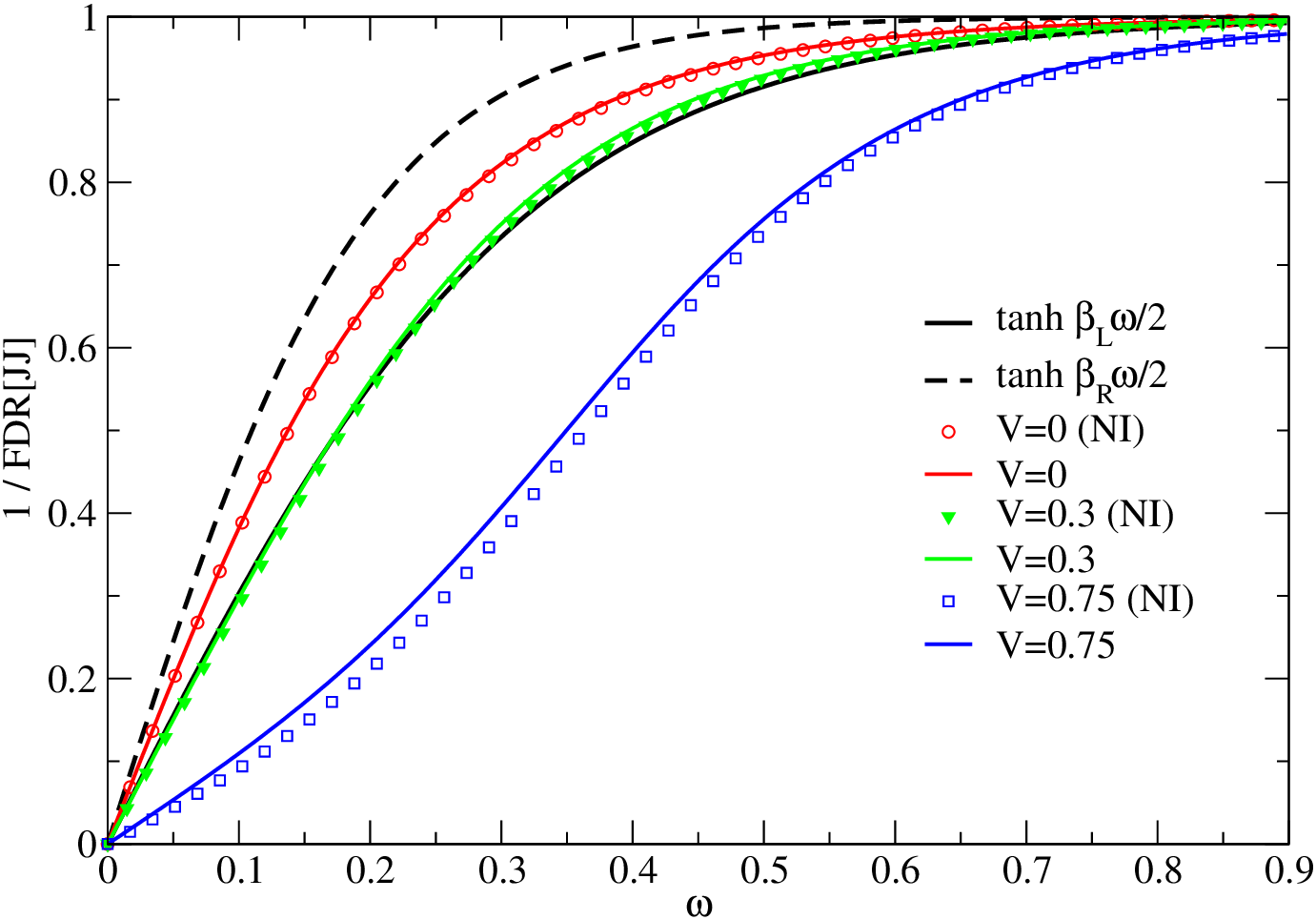} \\
  \includegraphics[clip,width=\mycolumnwidth,height=55mm]{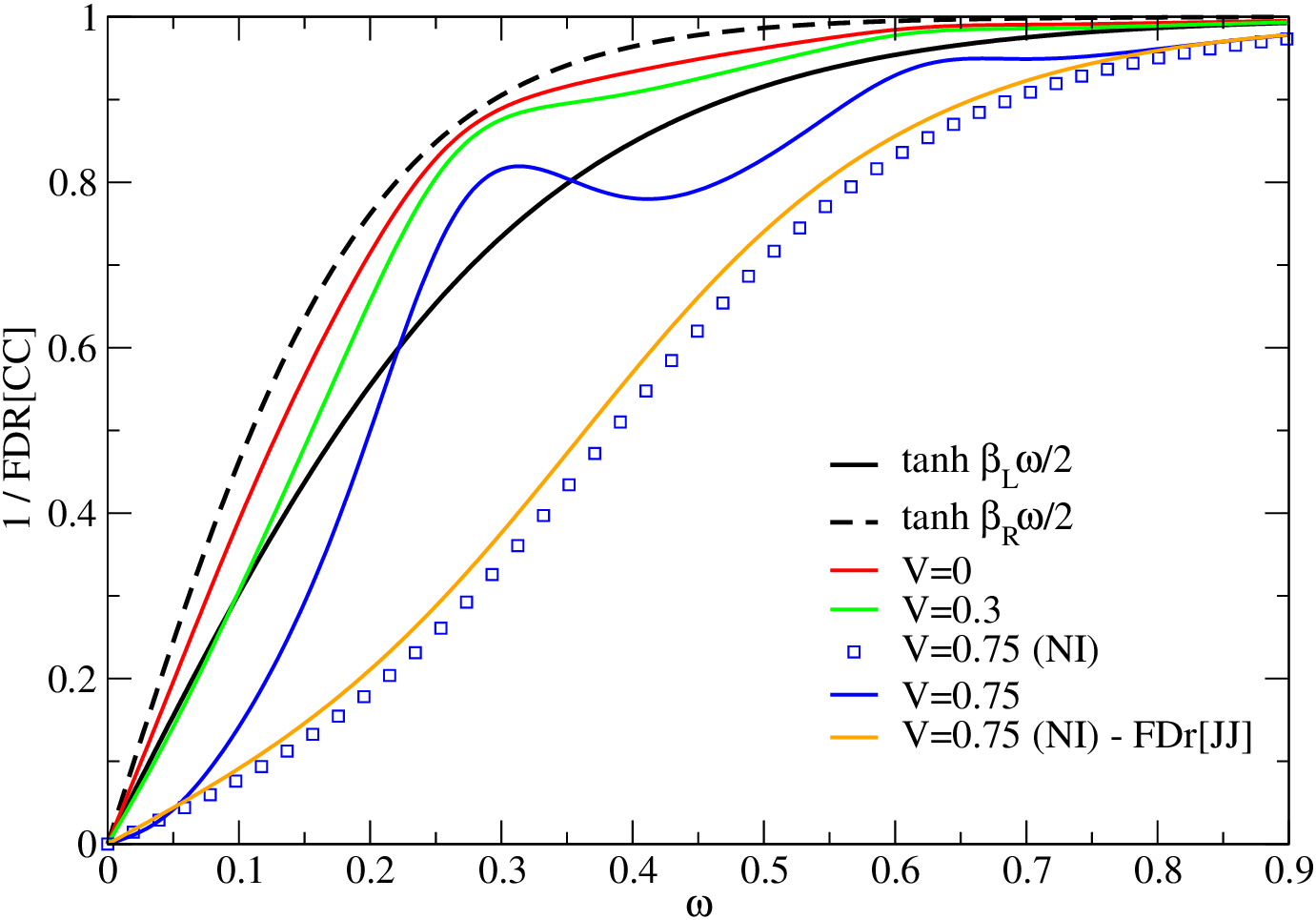}
  \caption{(color online)
FD ratio of the GFs (top panel) and of the JJ correlation functions
(middle panel) and the CC correlation functions (bottom panel) 
in the resonant transport regime ($\varepsilon_0=0.0$). Calculations are 
performed without the Hartree self-energy to preserve the full electron-hole symmetry
in the system.
The other parameters are the same of in Figure \ref{fig:9}.}
  \label{fig:9b}
\end{figure}

\section{Thermoelectric resonant transport regime}
\label{app:interac_thermoelec}

Figure \ref{fig:9b} shows the FDR of the GFs and the JJ and CC correlation functions
calculated in the resonant transport regime, in the presence of an applied bias and a
gradient of temperatures. By using only the Fock-like self-energy 
in the calculations, the system is always electron-hole symmetric.
Once again, one can see the dependence of the FDR[GF] and of the FDR[CC] on the NE and 
MB effects.
The effects of the interaction are still less pronounced in the FDR of the current-current
correlation functions.
Note that, in opposition to the off-resonant regime, the inverse FDR of the CC correlations
and the FDR of the GF  are qualitatively similar: they both present a oscillations with 
the energy argument.

We have also included the FDR[JJ] in the bottom panel of Figure \ref{fig:9b}
to illustrate once more that the FDR are different for the two different JJ and CC correlation
functions.

\end{document}